\PassOptionsToPackage{bookmarks=false}{hyperref}
\documentclass[10pt, journal, compsoc, hidelinks]{IEEEtran}
\usepackage{booktabs} 
\usepackage[utf8]{inputenc}
\usepackage[T1]{fontenc}
\usepackage{framed}
\usepackage{graphicx}
\usepackage{pifont}
\usepackage{xspace}
\usepackage{multirow}
\usepackage{xcolor,colortbl}
\usepackage{hhline}
\usepackage{adjustbox}
\usepackage{booktabs} 
\usepackage{amssymb}
\usepackage{balance}
\usepackage{tcolorbox}
\usepackage{url}
\usepackage{booktabs}
\usepackage{paralist}
\usepackage{enumitem}
\usepackage{pgfkeys}
\usepackage{lipsum}
\usepackage{hyperref}
\usepackage{framed}
\usepackage{changepage}
\usepackage[ruled,vlined,linesnumbered]{algorithm2e}

\usepackage[numbers,sort&compress]{natbib} 

\ifCLASSOPTIONcompsoc \usepackage[caption=false]{subfig}
\else
\usepackage[caption=false]{subfi g}
\fi

\usepackage{listings}
\usepackage[xindy]{glossaries}

\usepackage{cleveref}
\usepackage{tcolorbox}
\crefformat{Section}{\S~#2#1#3} 
\crefformat{subsection}{\S~#2#1#3}
\crefformat{subsubsection}{\S~#2#1#3}

\definecolor{codegreen}{rgb}{0,0.6,0}
\definecolor{codegray}{rgb}{0.73,0.38,0.06}
\definecolor{codepurple}{rgb}{0.27,0.38,0.97}
\definecolor{codemagenta}{rgb}{0.74,0.09,0.42}
\definecolor{backcolour}{rgb}{0.96,0.96,0.96}
\definecolor{ao}{rgb}{0.0, 0.0, 1.0}
\definecolor{ao(english)}{rgb}{0.0, 0.5, 0.0}

\newcommand*{\ColorIfNotInString}[1]{\color{codegray}\textbf{#1}}%
\newsavebox{\captionbox}
\sbox{\captionbox}{\parbox[t]{\linewidth}{A very, very long caption here}}
\lstdefinestyle{mystyle}{
    backgroundcolor=\color{backcolour},   
    commentstyle=\color{codegreen},
    keywordstyle=\color{ao},
    numberstyle=\tiny\color{codegray},
    stringstyle=\color{codemagenta},
    language=Java,
    breakatwhitespace=false,         
    breaklines=true,                 
    keepspaces=true,                 
    numbers=right,                    
    numbersep=5pt,
    xrightmargin=1.5em,               
    showspaces=false,                
    showstringspaces=false,
    showtabs=false,       
    captionpos=b,           
    tabsize=2,
    frame=tb,
    literate=%
    {0}{{{\ColorIfNotInString{0}}}}1
    {1}{{{\ColorIfNotInString{1}}}}1
    {2}{{{\ColorIfNotInString{2}}}}1
    {3}{{{\ColorIfNotInString{3}}}}1
    {4}{{{\ColorIfNotInString{4}}}}1
    {5}{{{\ColorIfNotInString{5}}}}1
    {6}{{{\ColorIfNotInString{6}}}}1
    {7}{{{\ColorIfNotInString{7}}}}1
    {8}{{{\ColorIfNotInString{8}}}}1
    {9}{{{\ColorIfNotInString{9}}}}1
}
\newcommand{\citeTodo}[1]{{\color{red}[??]}}

\newcommand{\ie}{\emph{i.e.},\xspace}

\newcommand{\eg}{\emph{e.g.},\xspace}

\newcommand{\etal}{\emph{et~al.}\xspace}

{

\newcommand\RQ[1]{\textbf{RQ$_#1$}}

\definecolor{gray50}{gray}{.5}
\definecolor{gray40}{gray}{.6}
\definecolor{gray30}{gray}{.7}
\definecolor{gray20}{gray}{.8}
\definecolor{gray10}{gray}{.9}
\definecolor{gray05}{gray}{.95}

\newboolean{showchanges}
\setboolean{showchanges}{true}
\ifthenelse{\boolean{showchanges}}
{

}
{   
	
}

\definecolor{highest}{HTML}{339999}
\newcommand{\researchquestion}[2]{ \vspace{-0.25 cm}
    \def\FrameCommand{%
        \hspace{3pt}%
        {\color{highest}\vrule width 3.0pt}%
        {\color{white}\vrule width 4pt}%
        \colorbox{white}%
    }%
    \MakeFramed{\advance\hsize-\width\FrameRestore}%
    \noindent\hspace{-4.55pt}
    \begin{adjustwidth}{}{-5pt}%
    
        \textbf{RQ$_#1$:} {#2}
    \end{adjustwidth}\vspace{-0.6pt}
    \endMakeFramed%
    \vspace{-0.25 cm} 
}

\newcommand{\resultsummary}[2]{ \vspace{-0.15 cm}
    \def\FrameCommand{%
        \hspace{3pt}%
        {\color{highest}\vrule width 2.5pt}%
        {\color{white}\vrule width 4pt}%
        \colorbox{white}%
    }%
    \MakeFramed{\advance\hsize-\width\FrameRestore}%
    \noindent \hspace{-4.55pt}
    \begin{adjustwidth}{}{-5pt}%
    	#1 \end{adjustwidth}\endMakeFramed%
    \vspace{-0.15 cm} 
}

\newlength\Linewidth
\def\findlength{\setlength\Linewidth\linewidth
  \addtolength\Linewidth{-4\fboxrule}
  \addtolength\Linewidth{-3\fboxsep}
}
\newenvironment{examplebox}{\par\begingroup
  \setlength{\fboxsep}{5pt}\findlength
  \setbox0=\vbox\bgroup\noindent
  \hsize=0.95\linewidth
  \begin{minipage}{0.95\linewidth}\normalsize}
  {\end{minipage}\egroup
  \textcolor{black}{\fboxsep1.5pt\fbox
    {\fboxsep5pt\colorbox{gray05}{\normalcolor\box0}}}
  \endgroup\par\noindent
  \normalcolor\ignorespacesafterend}

\newcommand\evosuite{\textsc{EvoSuite}\xspace}
\newcommand\mosa{MOSA\xspace}
\newcommand\pdmosa{aDynaMOSA\xspace}

\newcommand\dmosa{DynaMOSA\xspace}

\newcommand\rqone{(Effectiveness) What is the target coverage achieved by \pdmosa compared to \dmosa?}
\newcommand\rqoneshort{Effectiveness}

\newcommand\rqtwo{(Fault Detection) What is the mutation score achieved by \pdmosa compared to \dmosa?}
\newcommand\rqtwoshort{Fault Detection}

\newcommand\rqthree{(Performance) Does \pdmosa help to reduce test runtimes and heap memory consumption?}
\newcommand\rqthreeshort{Performance}

\newacronym{gc}{GC}{garbage collector}
\newacronym{vm}{VM}{virtual machine}
\newacronym{jvm}{JVM}{Java Virtual Machine}

\clubpenalty = 10000
\widowpenalty = 10000
\displaywidowpenalty = 10000
\begin{document}

\title{Testing with Fewer Resources:\\An Adaptive Approach to Performance-Aware Test Case Generation}

\author{Giovanni~Grano,
		Christoph~Laaber,
        Annibale~Panichella, and
        Sebastiano~Panichella
\IEEEcompsocitemizethanks{\IEEEcompsocthanksitem 
G. Grano and C. Laaber are with the University of Zurich, Switzerland
\protect\\
E-mail: \{\textit{grano, laaber}\}@ifi.uzh.ch
\IEEEcompsocthanksitem 
A. Panichella is with the Delft University of Technology, the Netherlands
\protect\\
E-mail: A.Panichella@tudelft.nl
\IEEEcompsocthanksitem 
S. Panichella is with the Zurich University of Applied Science, Switzerland. 
\protect
E-mail: panc@zhaw.ch
}
}

\IEEEcompsoctitleabstractindextext{
\begin{abstract} 
Automated test case generation is an effective technique to yield high-coverage test suites. 
While the majority of research effort has been devoted to satisfying coverage criteria, a recent trend emerged towards optimizing other non-coverage aspects.
In this regard, runtime and memory usage are two essential dimensions: less expensive tests reduce the resource demands for the generation process and later regression testing phases.
This study shows that \emph{performance-aware} test case generation requires solving two main challenges:
providing a good approximation of resource usage with minimal overhead and  
avoiding detrimental effects on both final coverage and fault detection effectiveness. 
To tackle these challenges, we conceived a set of performance proxies ---inspired by previous work on performance testing--- that provide a reasonable estimation of the test execution costs (\ie runtime and memory usage).
Thus, we propose an adaptive strategy, called aDynaMOSA, which leverages these proxies by extending DynaMOSA, a state-of-the-art evolutionary algorithm in unit testing.
Our empirical study ---involving 110 non-trivial Java classes--- reveals 
that our adaptive approach generates test suite with statistically significant improvements in runtime (-25\%) and heap memory consumption (-15\%) compared to DynaMOSA. Additionally, aDynaMOSA has comparable results to DynaMOSA over seven different coverage criteria and similar fault detection effectiveness.
Our empirical investigation also highlights that the usage of performance proxies (i.e., without the adaptiveness) is not sufficient to generate more performant test cases without compromising the overall coverage. 
\end{abstract}

}

\maketitle

\section{Introduction}
\label{sec:intro}
From \textit{Waterfall} to \textit{Agile}, software testing has always played an essential role in delivering high-quality software~\cite{fowler2006continuous}.
Integrating automated test case generation tools \cite{fraser2013whole, panichella15reformulating} in software development pipelines 
(\eg in continuous software development (CD) \cite{Vassallo2016}) could
potentially reduce the time spent by developers writing test cases by hand~\cite{campos2014continuous}.
Hence, research and industry have heavily focused on automated test generation in the last decade~\cite{mcminn2011search},
mainly employing evolutionary search (\eg genetic algorithms (GA)) to produce minimal test suites that satisfy some testing criteria~\cite{mcminn2004search-based}. 

While most of the research effort has been devoted to maximizing various code coverage criteria \cite{mcminn2011search,campos2017empirical,fraser2011evosuite,fraser2013whole}, 
recent work showed that further factors need to be considered for the generation of test cases \cite{lakhotia2007a,Afshan:icst2013,Xuan:fse2014,panichella2016impact}. 
Specifically, recent research investigated additional factors such as data input readability~\cite{Afshan:icst2013}, 
test readability~\cite{Ermira2015, panichella2016impact}, 
test code quality~\cite{palomba2016automatic}, 
test diversity~\cite{Albunian:ssbse17}, 
execution time~\cite{Xuan:fse2014},
and memory usage~\cite{lakhotia2007a}.
An early attempt to reduce the resource demand of generated tests is the work by Lakhotia \etal \cite{lakhotia2007a}.
The authors recast test data generation as a bi-objective problem where branch coverage and the number of bytes allocated in the memory are two contrasting objectives to optimize with Pareto-efficient approaches.
Their results show that multi-objective evolutionary algorithms are suitable for this problem. 
Following this line of research, other works also used multi-objective search to minimize test execution time \cite{pinto2010multi} 
or the number of generated tests, used as a proxy for the oracle cost \cite{oster2006automatic,ferrer2012evolutionary}.

While the aforementioned works showed the feasibility of lowering the cost (\eg execution time) of the generated tests,
they all pose \textit{two significant challenges} on the full code coverage \cite{ferrer2012evolutionary}.
The first challenge stems from empirical results showing that combining coverage with non-coverage criteria is harmful to the final coverage compared to traditional strategies that target coverage only~\cite{lakhotia2007a,oster2006automatic,ferrer2012evolutionary,palomba2016automatic}.
These approaches implement the classic \textit{one-branch-at-a-time} (or \textit{single-target}) approach, which consists of running bi-objective meta-heuristics (\eg GA) multiple times, once for every code branch, while performance aspects are other dimensions to optimize for each branch separately.
However, recent studies \cite{rojas2017detailed, panichella:ist2018, campos:ist2018} empirically and theoretically showed that single-target approaches are less effective and efficient than multi-target approaches (e.g., the whole suite approaches and many-objective search) in maximizing coverage. 
The second challenge to address regards how to inject test performance analysis into the main loop of multi-target strategies without incurring in a penalizing overhead.

Generated tests with lower resource demand might decrease the cost of introducing test case generation 
into continuous integration (CI) pipelines.
Hilton \etal~\cite{hilton2017trade} showed that acquiring hardware resources for self-hosted CI infrastructure is one of the main barriers for small companies when implementing CI policies:
more performant tests would require fewer hardware resources, and therefore testing in CI would be more cost-effective.
Despite the theoretical benefits, the precise measurement of memory and execution time costs adds considerable
overhead since it requires running each test case multiple times~\cite{bukh1992art}.
Consequently, there is a \textit{need for approaches that minimize the test resource demand \cite{ferrer2012evolutionary} without penalizing the final coverage nor the fault detection capability of generated tests}.

We extend the current state-of-the-art by proposing a novel \textit{adaptive} approach, called  \pdmosa (\textbf{A}daptive \textbf{\dmosa}), to address the two challenges described above. 
In designing our approach, we focus on  
(i) test execution time (runtime from now on), 
(ii) memory usage, 
(iii) code coverage, 
and (iv) fault detection capability as four relevant testing criteria in white-box test case generation.
To tackle the second challenge, we explored recent studies in performance testing \cite{de2017perphecy} and symbolic execution \cite{albert2011resource} that investigate suitable approaches to estimate the computational/resource demands of test cases.
In particular, we adopted three performance proxies ---computable with low overhead--- introduced by Albert~\etal~\cite{albert2011resource} for symbolic execution. 
Besides, we developed four additional performance proxies that provide an indirect approximation of the test execution costs (\ie runtime and memory usage).
These proxies, obtained through instrumentation, measure static and dynamic aspects related to resource usage through a single test execution: 
the number of objects instantiated (for heap memory), triggered method calls, and executed loop cycles and statements (for runtime).

Recent work in the field explored alternative ways to integrate orthogonal objectives into the fitness function, which are based on the idea of using non-coverage aspects as a second-tier objective \cite{palomba2016automatic}. 
To address our first challenge, \pdmosa extends \dmosa~\cite{panichella2018automated} ---the most recent many-objective genetic algorithm for test case generation--- by using the performance proxies as second-tier objectives while code branches are the first-tier objectives.
\pdmosa uses an adaptive strategy where the second objective \emph{can} be temporarily disabled in favor of achieving higher coverage values (which is the primary goal).  
We integrated an adaptive strategy in \pdmosa since our initial results show that when the second objective strongly competes with the primary one (\ie coverage), which is the case for performance, an adaptive strategy is preferable to a non-adaptive approach \cite{palomba2016automatic}.

To evaluate \pdmosa, we conduct an empirical study with 110 non-trivial classes from 27 open-source Java libraries
to show the usefulness of \pdmosa compared to the baseline \dmosa
in terms of branch coverage, runtime, memory consumption, and fault-effectiveness (\ie mutation score).
Our study shows that the test suites produced with \pdmosa are significantly less expensive to run for 72\% (for runtime) and 70\% (for heap memory consumption) of the classes compared to \dmosa. Besides, \pdmosa achieves similar code coverage compared to \dmosa over seven different testing criteria.
We demonstrate that the devised approach does not reduce the fault-effectiveness of the generated tests: 
\pdmosa achieves a similar or higher mutation score for \textasciitilde75\% of the subjects under tests.

\noindent\textbf{Contributions}.
We summarize the main contributions of this work in the following:
\begin{itemize}
	\item We propose a \emph{performance-score} aggregating a set of performance proxies with low overhead as an indirect approximation of the computational demand for a generated test case.
	\item We devise an \emph{adaptive} test case generation technique that leverage such performance proxies to optimize secondary objectives orthogonal to code coverage, without any negative effect on the latter.
	\item We instantiate our approach to the problem of reducing the \textit{resource} demand of generated test suites while maintaining high test coverage and fault detection capability.
	\item We show that the usage of performance proxies is not sufficient to achieve the best results. The key aspect of aDynaMOSA is its adaptive mechanisms that dynamically enables or disables the second objective depending on whether search stagnation of the coverage criteria is detected or not.
\end{itemize}
\noindent\textbf{Replication package}.
To enable full replicability of this study, we publish 
all the data used to compute the results and a runnable version of the implemented approach
in a replication package~\cite{appendix}.

\section{Background \& Related Work}
\label{sec:related}
Last decade witnessed extensive research on test data generation~\cite{mcminn2004search-based, 
mcminn2011search} aiming at generating tests with high code coverage, measured according to various code coverage criteria such as branch~\cite{Tonella:2004}, statement~\cite{mcminn2004search-based}, 
line, and method~\cite{campos2017empirical} coverage. 
Search-based algorithms ---GAs in particular \cite{goldberg2006genetic}--- had a strong pull on the automation of such a task \cite{mcminn2011search}.

Proposed approaches can be categorized into two formulations: \textit{single-target} and \textit{multi-target}. 
In the former, evolutionary algorithms (or more general 
meta-heuristics) aim to optimize one single-coverage target (e.g., branch) at one time. The single 
target $b$ is converted into a single function (\emph{fitness function}) measuring how 
close a test case (or a test suite) is to cover $b$~\cite{mcminn2004search-based}. 
The ``\emph{closeness}'' to a given branch is measured using two white-box heuristics~\cite{mcminn2004search-based}: the \textit{approach level} and the 
normalized \textit{branch distance}.
Fraser and Arcuri were the first to propose a multi-target approach, which optimizes all coverage targets 
simultaneously in order to overcome the disadvantages of targeting one branch at a 
time~\cite{fraser2013whole}.  
In their approach, called \textit{whole test suite generation} (WS), GAs evolve
entire test suites rather than single test cases. The search is then guided by 
a suite-level fitness function that sums up the coverage heuristics (\ie branch distances) for all the 
branches of the class under test (CUT). 
A later improvement over WS, called \textit{archive based whole suite 
approach} (WSA), focuses the search on uncovered branches only and uses an archive to collect test cases 
reaching uncovered branches~\cite{rojas2017detailed}.

%
%
\textbf{Many-objective search}.
Following the idea of targeting all branches at once, Panichella~\etal~\cite{panichella15reformulating}
addressed the test case generation problem in a many-objective fashion proposing a many-objective genetic 
algorithm called MOSA.
Different from WS (or WSA), MOSA evolves test cases that are evaluated using the 
\textit{branch distance} and \textit{approach level} for every single branch in the CUT.
Consequently, the overall fitness of a test case is measured based on a vector of $n$ objectives, one for 
each branch of the production code.
Thus, the goal is finding test cases that separately satisfy the target 
branches~\cite{panichella15reformulating}, \ie tests $T$ having a fitness score $f_i(T) = 0$ for at least 
one uncovered branch $b_i$.
To focus/increase the selection towards tests reaching uncovered branches, MOSA proposes a new way 
to rank candidate test cases \cite{von2014survey}, called \textit{preference criterion}.
Formally, a test case $x$ is \emph{preferred} over another test $y$ for a given branch $b_i$
(or $x \prec_{b_i} y$) \textit{iff} $f_i(x) < f_i(y)$~\cite{panichella15reformulating}, \ie its objective score is lower 
(main criterion). In addition, if the two test cases $x$ and $y$ are equally good in terms of branch 
distance and approach level for the branch $b_i$ (\ie $f_i(x) = f_i(y)$), the shorter test is preferred 
(secondary criterion). 
In other words, the preference criterion promotes test cases that are closer to cover some branches if 
possible and have minimal length.

MOSA works as follows\footnote{See Algorithm 1 in~\cite{panichella15reformulating} for full detail.}:
a starting \emph{population} is randomly generated and evolved through some \emph{generations}.
For each generation, new \emph{offsprings} are created through \emph{crossover} and \emph{mutation}.
Then, the new population for the next generation is created by selecting tests among parents and offsprings as 
follows: a first front $\mathbb{F}_0$ of test cases is built by using the \emph{preference criterion}.
Following, the remaining tests are grouped in subsequent fronts using the traditional \textit{non-dominated sorting 
algorithm}~\cite{aravind2004fast}. 
The new population is then obtained by picking tests starting from the first front  
$\mathbb{F}_0$ until reaching a fixed population size $M$. To enable diversity and avoid premature 
convergence \cite{kifetew2013orthogonal,Albunian:ssbse17}, MOSA also relies on the \emph{crowding distance}, 
a secondary heuristic that increases the chances to survive in the next generation for test cases that are 
the most diverse within the same front. The final test suite is the \emph{archive}, an 
additional data structure that stores test cases that reach previously uncovered branches.
If a new test $t$ hits an already covered branch $b_i$, $t$ is stored in the archive if and only if shorter (secondary criterion) than the test case stored in the archive for the same branch $b_i$.


Panichella \etal~\cite{panichella2018automated} improved the \mosa algorithm by 
presenting DynaMOSA.
Relying on the \emph{control dependency graph} (CDG), DynaMOSA narrows the search towards the uncovered targets free of control dependencies.
New targets are then iteratively considered when their dominators are satisfied.
In particular, the difference between \dmosa and \mosa is the following:
at the beginning of the search, DynaMOSA tries to hit only the targets free of any control dependencies.
Therefore, at every iteration, the current set of targets $U^*$ is updated based on the execution results of the newly generated offsprings: being $u_i$ a newly hit target, the targets dominated by $u_i$ are added to $U^*$.  
This approach does not impact the way \mosa ranks the generated solutions, but rather speeds up the convergence of the algorithm, while optimizing the size of the current objects.  
Empirical results show that DynaMOSA performs better than both WSA 
and MOSA in terms of branch~\cite{panichella2018automated, campos2017empirical}, 
statement~\cite{panichella2018automated}, and strong mutation coverage~\cite{panichella2018automated}.

Recently, Panichella \etal~\cite{panichella2018incremental} further improved DynaMOSA with the goal of maximizing different coverage criteria simultaneously (\eg branch, line, and weak mutation coverage, at the same time).
The latest variant of DynaMOSA relies on the \emph{enhanced control dependency graph} (ECDG) enriched with structural dependencies among the testing targets.
These objectives to optimize are dynamically selected using the ECDG while exploring the covered control dependency frontier incrementally.
Empirical results show that even though the multi-criteria variant may result in few cases in a lower branch coverage than \dmosa, it reaches higher coverage on all the other criteria as well as showing better fault detection capability~\cite{panichella2018incremental}.
We use this many-criteria version of \dmosa both as a baseline and a starting point for implementing the proposed adaptive approach.

\textbf{Large-scale studies}.
Campos \etal~\cite{campos:ist2018} and Panichella \etal~\cite{panichella:ist2018} conducted two large-scale empirical studies comparing different approaches and meta-heuristics for test case generation. Their results showed that: (1) multi-target approaches are superior to the single-target approaches, and (2) many-objective search helps to reach higher coverage than alternative multi-target approaches for a large number of classes. Besides, no search algorithm is the best for all classes under test~\cite{campos:ist2018}. These recent advances motivate our choice of focusing on many-objective search.

\textbf{Non-coverage objectives}. In recent years, several works focused on \textit{non-coverage} aspects in addition to reaching high coverage.
Lakhotia~\etal proposed a multi-objective formulation optimizing branch coverage as primary and dynamic memory consumption as secondary objective~\cite{lakhotia2007a}.
Ferrer~\etal proposed a multi-objective approach aiming at simultaneously maximizing code coverage and minimizing oracle cost~\cite{ferrer2012evolutionary}. 
Afshan~\etal focused on code readability as a crucial secondary objective to foster maintainability~\cite{afshan2013evolving}.
In particular, their approach generates readable string inputs exploiting natural language models.
Despite empirical research showed the difficulty of effectively balancing two contrasting objectives without penalizing the final code coverage~\cite{ferrer2012evolutionary}, the mentioned studies all gave the same weight to coverage and non-coverage objectives.
Furthermore, these studies used a single-target approach rather than multi-target ones.
Palomba \etal~\cite{palomba2016automatic} incorporated test cohesion and coupling metrics as secondary objectives within the \textit{preference criterion} of MOSA to produce more maintainable test cases, from a developer point of view. 
Their approach produces more-cohesive and less-coupled test cases without reducing coverage. 
More recently, Albunian~\cite{Albunian:ssbse17} investigated test case diversity as a further objective to optimize together with coverage in WSA. 

\textbf{Our work}.
We propose aDynaMOSA, a novel test case generation algorithm that optimizes a secondary objective besides code coverage.
Differently from most previous attempts of combining non-coverage with coverage objectives, aDynaMOSA relies on many-objective search.
To balance the two orthogonal objectives, it adaptively enables or disables the optimization of the secondary objective when adverse effects on the code coverage are detected during the generation.
In this work, we instantiate aDynaMOSA to focus on the performance ---\ie runtime and heap memory consumption--- of generated tests.
To achieve this goal, we utilize metrics approximating test case performance while having low analysis overhead (\cref{sec:indicators}).

\section{Approach}
\label{sec:approach}
This section introduces the performance proxies, their rationale, and how we integrated them in \dmosa. 

\subsection{Performance Proxies}
\label{sec:indicators}
The accurate measurement of software system performance is known to be challenging:
it requires measurements to be performed over multiple runs to account for run-to-run variations~\cite{bukh1992art}.
This means that we would need to re-run each generated test case hundreds of times to have rigorous runtimes and memory usages.
This type of direct measurement is unfeasible for test case generation,
where each search iteration generates several new tests that are typically executed only once for coverage analysis. 

While a direct measurement is unfeasible in our context, various test case characteristics can be used to indirectly estimate the cost (runtime and memory) of the generated tests.
According to Jin~\etal \cite{jin2012}, about 40\% of real-world performance bugs stem from inefficient loops, 
while uncoordinated method calls and skippable functions account for respectively one third and a quarter of performance bugs.
Object instantiations impact the \textit{heap memory} usage~\cite{shirazi2003java},
and the number of executed statements has been used in previous regression testing studies as a proxy for runtime~\cite{Yoo:2007}.
Multiple studies investigate the performance impact prediction in the context of software performance analysis~\cite{huang2014, de2017perphecy, mostafa2017},
but to the best of our knowledge, no prior work combined it with evolutionary unit test generation.

The closest studies are the ones from De~Oliveira \etal~\cite{de2017perphecy} and Albert \etal~\cite{albert2011resource}, which fit the context of this study. 
However, both studies leveraged only a subset of proxies investigated in this paper and focused on different testing problems and techniques. De~Oliveira~\etal~\cite{de2017perphecy} investigated performance proxies in the context of regression testing. 
Albert~\etal~\cite{albert2011resource} proposed three performance proxies for symbolic execution and showed their benefits on example programs. 
Symbolic execution can be used as an alternative technique to generate test cases rather than GAs; however, it has various limitations widely discussed in the literature~\cite{chen2015star,soltani2018search}, such as the path explosion problem, it cannot handle external environmental dependencies, and complex objects.

In this paper, we extend the set of performance proxies proposed in previous studies~\cite{albert2011resource} and incorporate them within evolutionary test case generators in an adaptive fashion. We designed the performance proxies 
with the idea of estimating a test  case's performance (\ie runtime and/or memory consumption) unobtrusively.
We implemented two types of proxies: 
(i) \emph{Static proxies} that utilize static analysis techniques such as AST parsing. 
(ii) \emph{Dynamic} proxies that rely on the instrumentation facilities available in \textsc{EvoSuite}~\cite{fraser2011evosuite}.
We extract the \textit{control flow graph} (CFG) and the number of times each branch in the CFG is covered by a given test $t$ (frequency).
All production code proxies are dynamic while proxies related to the test code are static.

Table~\ref{table:proxies} summarizes the performance proxies.
In the following, we describe them separately and discuss which dimension (memory or runtime) they affect.

\textbf{Number of executed loops} (\textbf{I$_1$}). 
This counts the number of loop cycles in the production code which are executed/covered by a given test case $t$.
Higher loop cycle counts influence the runtime of the test case.
To this aim, at instrumentation time, we use a \textit{depth-first traversal} algorithm to detect loops in the CFG. 
When a test case $t$ is executed, we collect the number of times each 
branch involved in a loop is executed (execution frequency). 
Thus, the proxy value  
for $t$ corresponds to the sum of the execution frequencies for all branches involved in loops. 
To avoid a negative impact on coverage, we require each loop to be covered at least once.
Therefore, this proxy only considers loops with a frequency higher than one.

\textbf{Number of method calls}~\cite{albert2011resource}.
We implement two types of method call proxies:
\textbf{covered method calls} (\textbf{I$_2$}), which is related to method calls in the paths of 
the CFG that are covered by a test $t$; 
and \textbf{test case method calls} (\textbf{I$_3$}), which counts method 
calls in $t$. 
Notice, the former proxy considers the number of calls to each production method (\ie the frequency) 
rather than a single boolean value denoting whether a method has been called or not, as in method coverage~\cite{campos2017empirical}.
This is because a method can be invoked multiple times by either indirect calls or within loops. 
Method calls directly impact the memory usage~\cite{McAllister:2008}: every time a method is invoked, a new frame is allocated on top of the \textit{runtime stack}.
Further, method calls are dynamically dispatched to the right class, which might influence the runtime.
Thus, fewer method calls should result in shorter runtimes and lower heap memory usage due to potentially fewer object instantiations.

\textbf{Number of object instantiations} (\textbf{I$_4$}).
Objects instantiated during test executions are stored on the \textit{heap}. 
Thus, reducing the number of instantiated objects may lead to decreased usage of heap memory. 
The fourth proxy counts the number of object instantiations triggered by a test case 
$t$. 
It analyzes the basic blocks of the CFG that $t$ covers and 
increments a counter for every constructor call and local array definition statement.
Notice that we consider the frequency (\eg the number of constructor calls) rather than a binary value (\ie called or not called).
Moreover, the constructor call counter excludes calls and local array definitions with a frequency of one, as we want to cover them at least once. 
We do not consider the instantiated-object size as it would require more complex and 
heavier instrumentation. 
We also do not consider primitive data types which use memory as well, because their 
influence is negligible compared to objects~\cite{shirazi2003java}.

\textbf{Number of statements}~\cite{albert2011resource}.
Statement execution frequency is a well-known proxy for runtime~\cite{Yoo:2007}.
Similarly to the proxies for number of method calls, we implement two types of statement-related proxies:
\textbf{Covered statements} (\textbf{I$_5$}), which counts the statements in the production code covered by a test case.
This proxy utilizes the dynamically-computed CFG for counting the covered statements.
\textbf{Test case statements} (\textbf{I$_6$}), which corresponds to the number of non-method-call statements in a generated test case.
This number is statically determined by inspecting the abstract syntax tree of the test case.

\textbf{Test case length (\textbf{I$_7$})}.
This counts the LOC (\emph{size}) of a test case and therefore represents a superset of test case method calls (\textbf{I$_7$}) and test case statements (\textbf{I$_6$}).
We include this proxy for two reasons:
First, it is a good performance proxy: longer tests mean more method and statement calls.
Second, \dmosa uses test case length as a secondary objective in order to reduce the oracle cost~\cite{barr2015oracle}.
Thus, we rely on the same metric to have a fair comparison.

\begin{table*}[thb]
\centering
\caption{Description of the performance proxies}
\label{table:proxies}
\resizebox{\linewidth}{!}{
\begin{tabular}{lll}
\toprule
ID & Performance Proxy & Description \\
\midrule
$I_1$ & Number of Executed Loops & Counts the number of loop cycles in the production code executed by a given test $t$ \\
$I_2$ & Covered Method Calls & Counts the number of times a method of the production code is called during the execution of a test $t$  \\
$I_3$ & Test Case Method Calls & Counts the number of methods calls contained in a test $t$ \\
$I_4$ & Number of Object Instantiations & Counts the number of objects (not primitive data type) instantiated during the execution of a test $t$\\
$I_5$ & Covered Statements & Counts the number of statements of the production code executed running a test $t$ \\
$I_6$ & Test Case Statements & Counts the number of non-method-call statements in a test $t$ \\
$I_7$ & Test Length & Counts the lines of code (LOC) of a test $t$\\
\bottomrule
\end{tabular}
}
\end{table*}

\subsection{Performance-Aware Test Case Generation}
\label{sec:algo}
To successfully generate test suites with high target coverage and, \emph{at the same time}, low computational 
requirements, we incorporate the performance proxies, described in \cref{sec:indicators}, into the 
main loop of \dmosa~\cite{panichella2018automated}. 
We opt for \dmosa since it has been shown to outperform other search algorithms (\eg WS, WSA, and MOSA) in branch and mutation coverage, positively affecting the test generation performance~\cite{panichella2018automated}.
Additionally, its basic algorithm (\ie \mosa) was used in prior studies to combine multiple testing criteria \cite{rojas2015combining, palomba2016automatic}.
Multiple approaches could be followed to this aim. 
One theoretical strategy consists of adding the performance proxies as further 
search objectives in addition to the coverage-based ones, merely following the many-objective
paradigm of \mosa. 
This leads to a trade-off search between coverage and 
non-coverage objectives that is not meaningful in testing~\cite{palomba2016automatic}.
Test cases that reduce the memory usage but at the same time reduce the final coverage are of less interest.
Therefore, considering coverage and non-coverage criteria as equally important objectives results in tests with decreased coverages~\cite{ferrer2012evolutionary, lakhotia2007a, pinto2010multi, 
oster2006automatic,Albunian:ssbse17}.

For these reasons, we investigate an alternative strategy where performance proxies 
are considered as \textit{secondary objectives} while coverage is the \textit{primary objective}. 
At first, we experiment with the most straightforward possible approach, \ie using the performance proxies as secondary criteria, as proposed in a prior study~\cite{palomba2016automatic}. 
However, due to the negative impact of the proxies on the final coverage, we refine this strategy by using an adaptive mechanism that enables and disables the proxy usage depending on whether \textit{search stagnation} is detected or not.
We refer to this adaptive strategy as \pdmosa (\textbf{A}daptive \textbf{\dmosa}) (\cref{section:pmosa}).
%
%
\subsubsection{Performance-Score as Secondary Objective}
\label{section:performance:secondary}
We first integrate the performance proxies using the methodology proposed in a prior study~\cite{palomba2016automatic}.
This approach replaces the original secondary criterion of MOSA (test case length) with a \emph{quality score} based on test method coupling and cohesion.
Therefore, it uses the new secondary criterion in two points:
(i) in the \emph{preference criterion} to build the first front $\mathbb{F}_0$, selecting the test case with the lowest \emph{quality score} in the case many of them have the same minimum object value for an uncovered branch $b_i$; and  
(ii) in the routine used to update the \textit{archive}.

In this first formulation, we adopt the same methodology replacing the \emph{quality score} with the \emph{performance-score} computed for each test case $t$ as follows:
\begin{equation}\label{eq:score}
\textit{performance-score}(t) = \sum_{I_k \in I} \frac{I_k(t)}{I_k(t)+1}
\end{equation}
where $I$ denotes the seven proxies described in \cref{sec:indicators}.
To deal with different magnitudes, each proxy value $I_k(t)$ is normalized in Equation~\ref{eq:score} using the normalization function $f[I_k(t)]=I_k(t)/[I_k(t)+1]$ \cite{mcminn2004search-based,arcuri:2010}.

A preliminary evaluation of this strategy showed that the performance proxies ---\ie even as a secondary criterion--- lead to a dramatic reduction of branch coverage.
We observed that the performance proxies strongly compete with coverage, \eg test cases that trigger fewer method calls likely lead to lower code coverage.
For this reason, we devise a second approach called \pdmosa that overcomes this limitation. 
We include the preliminary approach's results in the replication package \cite{appendix}.

%
%
\subsubsection{Adaptive DynaMOSA (\pdmosa)}
\label{section:pmosa}
\pdmosa
uses an adaptive mechanism to decide whether to (not) apply the performance proxies depending on the search improvements 
done during the generations. 
We devise this strategy because continuously selecting test cases with the lowest performance proxies value leads to reduced code coverage. 

The pseudo-code of \pdmosa is outlined in Algorithm~\ref{algo:adaptive:mosa}.
\begin{algorithm}[!tb]
  \DontPrintSemicolon
  \scriptsize
  \SetAlgoLined
  \SetAlgoVlined
  \SetKwInOut{Input}{Input}
  \Input{$B=\{\tau_1,...,\tau_m\}$: set of coverage targets of a program\\
  \indent Population size $M$\\
  \indent $CDG = \langle N, E, s\rangle$: control dependency graph of a program\\
  }
  \KwResult{A test suite $T$}
   \Begin{
		 $\phi \leftarrow$ EXTEND-CDG($CDG, B$)\;
  		 $i \leftarrow 0$\;
  		 $B^* \leftarrow $ ENTRY-POINTS($CDG, \phi$, $B$)\;
  		 $P_i \leftarrow$ RANDOM-POPULATION($M$)\;
  		 \textit{Archive} $\leftarrow$ \textbf{PERFORMANCE-UPDATE-ARCHIVE}($P_i$, $\emptyset$)\;
  		 $B^* \leftarrow$ UPDATE-TARGETS($B^*, CDG, \phi$)\;
  		 \While{not(search budget consumed) AND ($B \neq \emptyset$)}{
  			 $O_i \leftarrow$ GENERATE-OFFSPRING($P_i$)\;
  			 \textit{Archive} $\leftarrow$ \textbf{PERFORMANCE-UPDATE-ARCHIVE}($O_i$, \textit{Archive})\;
  			 $H_i \leftarrow$ \textbf{GET-SECONDARY-HEURISTIC}($O_i$, $i$)\\
  			 $B^* \leftarrow$ UPDATE-TARGETS($B^*, CDG, \phi$)\;
  			 $\mathbb{F} \leftarrow$ PREFERENCE-SORTING($P_i \bigcup O_i, B^*$)\;
  			 $P_{i+1} \leftarrow \emptyset$\;
  			 $d \leftarrow 0$\;
  			 \While{$|P_{i+1}| + |\mathbb{F}_d| \leq M$}{
  				 \eIf{$H_i$ is crowding-distance} {
  					 \textbf{CROWDING-DISTANCE-ASSIGNMENT}($\mathbb{F}_d$)\;	
  				}
  				{
  				 \textbf{PERFORMANCE-SCORE-ASSIGNMENT}($\mathbb{F}_d$)}
  				 $P_{i+1} \leftarrow P_{i+1} \bigcup \mathbb{F}_d$\;
  				 $d \leftarrow d+1$\;
  				 }
  			 Sort($\mathbb{F}_d$) \tcc*[l]{according to $H_i$}
  			 $P_{i+1} \leftarrow P_{i+1} \bigcup \mathbb{F}_d[1:(M- | P_{i+1} |)]$\;
  			 $i \leftarrow i+1$\;
  		}
  		 \Return{$T$} \;
  }
  \caption{\pdmosa Pseudo-Algorithm}
  \label{algo:adaptive:mosa}
\end{algorithm}

We highlight in bold the modifications over the original DynaMOSA algorithm.
\pdmosa first builds the ECDG (line 2 of Algorithm~\ref{algo:adaptive:mosa}) as done in DynaMOSA~\cite{panichella2018incremental}. The initial set of coverage targets (objectives) is selected using the ECDG~\cite{panichella2018incremental} (line 4).
Subsequently, an initial population of $M$ test cases is randomly generated (line 5), and the archive is updated by storing tests that reach previously uncovered targets (line 6).
In each iteration, \pdmosa updates the set of objectives to optimize based on to the test execution results~\cite{panichella2018automated, panichella2018incremental} (lines 6 and 10 of Algorithm~\ref{algo:adaptive:mosa}).
The while loop in lines 8-25 evolves the test cases until all the objectives are satisfied or the 
search budget is over~\cite{panichella2018incremental}. New test cases (\textit{offsprings}) are synthesized in line as done in \dmosa~\cite{panichella2018incremental}:
(i) selecting parents with a \textit{tournament selection}, 
(ii) combining parents with a \textit{single-point crossover}, and 
(iii) further mutating the generated offspring tests with the 
\textit{uniform mutation}. 
Newly generated tests are executed against the CUT~\cite{panichella2018incremental}. Besides, the
corresponding objective scores~\cite{panichella2018incremental} and performance proxy values are also computed at this stage. 

Next, parents and offsprings are ranked into non-dominance fronts (line 13) using the \textit{preference sorting algorithm} and \textit{non-dominated sorting algorithm} as done in MOSA~\cite{panichella15reformulating} and \dmosa~\cite{panichella2018automated}. The main difference between \pdmosa and \dmosa lies in the secondary heuristic used to rank parent and offspring tests.  In MOSA, the secondary heuristic is the \textit{crowding distance}, 
which promotes more diverse test cases within the same front. The 
\textit{crowding distance} is responsible for ensuring diversity among the selected tests~\cite{aravind2004fast}, which is a critical 
aspect of evolutionary algorithms~\cite{vcrepinvsek2013}. 
A lack of diversity leads to \textit{stagnation} in local 
optima~\cite{vcrepinvsek2013,panichella2018automated}, which could reduce the probability to cover feasible branches.

In \pdmosa, we use both the \textit{crowding distance} and the \textit{performance proxies} as secondary 
heuristics. 
\pdmosa uses the routine GET-SECONDARY-HEURISTIC (lines 11 and 17-20 of Algorithm 
\ref{algo:adaptive:mosa}) to decide which of the two alternative secondary heuristics to apply, which depends on whether search stagnation is detected or not. 
Algorithm~\ref{algo:check:stagnation} depicts the pseudo-code of the routine GET-SECONDARY-HEURISTIC. 
In the first generation, the default secondary 
heuristic is the one based on performance proxies (lines 2-5 of  Algorithm~\ref{algo:check:stagnation}). 
\begin{algorithm}[!bt]
  \scriptsize
  \DontPrintSemicolon
  \SetAlgoLined
  \SetAlgoVlined
  \SetKwInOut{Input}{Input}
  \Input{$Q_i$: new offsprings; $i$: the current iteration;\\
  \indent $B=\{\tau_1,...,\tau_m\}$: set of coverage targets of a program}
  \KwResult{$H_i$: heuristic for the current generation}
  \Begin{
     \If{i==0}{
      \tcp*[l]{Counters for generations with stagnation}
      performance-counter $\leftarrow$ 0 \\
      crowding-counter $\leftarrow$ 0\\
      \Return{performance-heuristic} \tcc*[r]{Initial heuristic}
     }
     stagnation $\leftarrow$ TRUE\\
       \For{$b \in B$ and $b$ is not covered}{
         \If{best objective value for $b$ in $Q_{i}$ better than in $Q_{i-1}$}{
          stagnation $\leftarrow$ FALSE\\
         }
       }
       \eIf{stagnation}{
        \tcp*[l]{Heuristic with the lowest stagnation counter}        
        \eIf{$H_{i-1}$ is performance-heuristic}{
          performance-counter $\leftarrow$ performance-counter+1\\
        } {
          crowding-counter $\leftarrow$ crowding-counter+1\\
        }
        \eIf{performance-counter $\leq$ crowding-counter}{
          \Return{performance-heuristic}
      } {
        \Return{crowding-distance}
      }
       }{
        \tcp*[l]{Heuristic used in the previous iterations}
        \eIf{$H_{i-1}$ is performance-heuristic}{
          performance-counter $\leftarrow$ 0\\
        } {
          crowding-counter $\leftarrow$ 0\\
        }
        \Return{$H_{i-1}$}
       }
    }   
  \caption{GET-SECONDARY-HEURISTIC}
  \label{algo:check:stagnation}
\end{algorithm}
For the later 
generations, the secondary heuristic is chosen by (i) analyzing the current objective scores to detect 
stagnation and (ii) taking into account which heuristics were used in the previous generations. 
Stagnation is detected when no improvement is observed for all uncovered branches (lines 7-9), \ie the fitness functions for all coverage criteria are unchanged over the last two generations.
Then, two counters are used to keep track of how often (\ie in how many iterations) stagnation was detected when either applying the \textit{crowding distance} or using the \textit{performance proxies}. 
In case of stagnation, the algorithm selects a new secondary heuristic with the lowest stagnation counter (lines 11-18 of Algorithm \ref{algo:check:stagnation}).
Otherwise, the secondary heuristic for the current generation $i$ remains 
the same as used in the previous iteration $i-1$ (lines 20-24).

Once the secondary heuristic for the current iteration is selected, \pdmosa assigns a secondary score to every test case in each dominance 
front $\mathbb{F}_d$ (lines 18 and 20 of Algorithm~\ref{algo:adaptive:mosa}) based on either the crowding distance or the
performance proxies.
If the employed secondary heuristic is the crowding distance, the 
secondary score of the tests corresponds to the crowding distance scores computed using the \textit{subvector
dominance assignment} by K\"{o}ppen et al.~\cite{Koppen:2007,panichella2018automated}.
Otherwise, if the performance proxies are selected for the 
secondary heuristic, the secondary score for each test case $t$ is computed as follows:
\begin{equation}
	\textit{performance-heuristic}(t)=\sum_{I_k \in I} \frac{I^{max}_{k}(\mathbb{F}_d) - I_k(t)}{I^{max}_{k}(\mathbb{F}_d) - I^{min}_{k}(\mathbb{F}_d)}
\end{equation}
where $I_k(t)$ is the value of the $k$-th proxy for the test $t$; $I^{max}_{k}(\mathbb{F}_d)$ and 
$I^{min}_{k}(\mathbb{F}_d)$ are the maximum and the minimum values of the $k$-th proxy 
among all tests in the front $\mathbb{F}_d$. 
The performance-heuristic takes a value in $[0;7]$; a zero value is 
obtained when the test case $t$ has the largest (worst) proxy values among all tests within the same front $
\mathbb{F}_d$, i.e., $\forall I_k, \;I_k(t)=I^{min}_{k}(\mathbb{F}_d) $; a maximum value of seven 
(corresponding to the total number of proxies) is obtained when $t$ has the lowest (best) proxies values 
among all tests within the same front $\mathbb{F}_d$, \ie 
$\forall I_k, \;I_k(t)=I^{max}_{k}(\mathbb{F}_d)$. Therefore, higher values of the performance-heuristic are 
preferable.

Crowding distance and performance-heuristic are then used in lines 21 
and 23 of Algorithm~\ref{algo:adaptive:mosa} to select test cases from the fronts 
$\mathbb{F}_0$-$\mathbb{F}_k$ until it reaches a maximum population size of $M$. 
When the \textit{crowding-distance} is used, more diverse tests within each front have a higher probability of being selected for the next population.
On the other hand, when the performance-heuristic is used, the tests with lower predicted resource demands are favored. 
Note, we update the archive based on the predicted performance of the executed test cases,
The update of the archive works as follows: 
when a test case $t$ satisfies an uncovered branch $b_i$, $t$ is automatically added to the archive.
Otherwise, if a new test $t$ hits an already covered branch $b_i$, $t$ is added to the archive if and only if its \textit{performance-score} (Equation~\ref{eq:score}) is lower than the score of the test case in the archive for $b_i$.
On the contrary, DynaMOSA employs the \textit{preference-criterion}.

\section{Empirical Study}
\label{sec:study}
Our empirical evaluation compares \pdmosa with \dmosa along three dimensions:
(i) seven different coverage criteria (the default ones of \evosuite),
(ii) fault detection effectiveness measured by strong mutation,
and (iii) resource usage measured by runtime and heap memory consumption.
Therefore, we investigate the following research questions:

\researchquestion{1}{\emph{\rqone}}

We evaluate the seven default criteria that are available in \evosuite optimized by \pdmosa via many-objective optimization.
The criteria are: branch, line, weak mutation, method, input, output, and exception coverage~\cite{rojas2015combining,panichella2018incremental}.
With \textbf{RQ$_1$}, we investigate whether and to what degree the introduction of the performance proxies affects the target coverage of each criterion.
\researchquestion{2}{\emph{\rqtwo}}

The second research question extends the comparison between \pdmosa and \dmosa by focusing on fault detection.
Tests generated using the proposed performance-aware approach might have a different structure (\eg contain fewer statements and method calls).
Therefore, we conduct a mutation-based analysis assessing whether optimizing performance proxies impact the fault detection capability of the generated tests.

\researchquestion{3}{\emph{\rqthree}}

The last research question investigates whether the approach is able to 
generate tests with reduced resource usage. In particular, we investigate two dimensions: \emph{time}, measuring runtime; and \emph{memory} looking at the heap memory consumption of the generated tests.

For both RQ1 and RQ2, we also compare our approach with random search,
which is a common baseline when using search-based techniques~\cite{arcuri2011practical}.

\medskip
\noindent\textbf{Prototype tool}.
\label{sec:study:tool}
We implemented \pdmosa in a prototype tool extending the \evosuite test suite generation framework,
as explained in \cref{sec:algo}.
The source code of the prototype tool is available on GitHub.\footnote{\url{https://github.com/giograno/evosuite}}
All experimental results reported in this paper are obtained using this prototype tool. 
Moreover, a runnable version of the tool itself is available for download in the replication package~\cite{appendix}.

\subsection{Subjects}
\label{sec:study:subjects}
\begin{table}[t]
\centering
\caption{Java Projects and Classes in Our Study}
\label{table:subjects}
\resizebox{\linewidth}{!}{
\begin{tabular}{@{}lrrrrrrr@{}}
\toprule
         Project & \# & \multicolumn{3}{c}{Branches} & \multicolumn{3}{c}{Mutants} \\
         \cmidrule(r){3-5} \cmidrule{6-8}
                 & ~     &            Min &   Max &  Mean &     Min &    Max &  Mean \\
\midrule
 a4j             & 2  & 30  & 124   & 77    & 15    & 911    & 463 \\
 bcel            & 4  & 52  & 890   & 475   & 408   & 1,523  & 1,043 \\
 byuic           & 1  & 722 & 722   & 722   & 2,173 & 2,173  & 2,173 \\
 fastjson        & 10 & 20  & 2,880 & 564   & 36    & 13,152 & 2,078 \\
 firebird        & 3  & 90  & 194   & 131   & 347   & 441    & 392 \\
 fixsuite        & 1  & 32  & 32    & 32    & 110   & 110    & 110 \\
 freehep         & 6  & 48  & 160   & 92    & 112   & 807    & 297 \\
 freemind        & 1  & 170 & 170   & 170   & 2,427 & 2,427  & 2,427 \\
 gson            & 4  & 60  & 660   & 285   & 126   & 2,870  & 1,212 \\
 image           & 7  & 34  & 274   & 140   & 214   & 1,676  & 589 \\
 javathena       & 1  & 230 & 230   & 230   & 752   & 752    & 752 \\
 javaviewcontrol & 2  & 212 & 2,360 & 1,286 & 2,058 & 4,972  & 3,515 \\
 jdbacl          & 2  & 170 & 174   & 172   & 595   & 700    & 648 \\
 jiprof          & 1  & 816 & 816   & 816   & 6,420 & 6,420  & 6,420 \\
 jmca            & 2  & 198 & 1,696 & 947   & 2,436 & 9,669  & 6,052 \\
 jsecurity       & 1  & 52  & 52    & 52    & 165   & 165    & 165 \\
 jxpath          & 3  & 98  & 102   & 100   & 204   & 449    & 312 \\
 la4j            & 7  & 20  & 280   & 135   & 196   & 3,217  & 1,122 \\
 math            & 4  & 14  & 238   & 92    & 135   & 1,274  & 443 \\
 okhttp          & 5  & 64  & 542   & 194   & 200   & 2,571  & 846 \\
 okio            & 9  & 24  & 562   & 126   & 34    & 4,271  & 1,009 \\
 re2j            & 8  & 68  & 646   & 178   & 148   & 2,096  & 1,129 \\
 saxpath         & 1  & 458 & 458   & 458   & 659   & 659    & 659 \\
 shop            & 4  & 38  & 182   & 102   & 175   & 465    & 302 \\
 webmagic        & 4  & 10  & 142   & 84    & 29    & 337    & 201 \\
 weka            & 10 & 212 & 778   & 359   & 255   & 13,263 & 2,220 \\
 wheelwebtool    & 7  & 24  & 804   & 331   & 75    & 3,898  & 1,637 \\
\midrule
Total & 110 & \multicolumn{6}{c}{~} \\
\bottomrule
\end{tabular}
}
\end{table}
Our benchmark consists of Java classes from different test benchmarks widely used in the SBST (Search-Based Software Testing) community:
(i) the SF110 corpus \cite{fraser2014large},
(ii) the 5\textsuperscript{th} edition of the Java Unit Testing Tool Competition at SBST 2017 \cite{panichella2017java},
and (iii) benchmarks used from previous papers about test data generation \cite{panichella2018automated, panichella15reformulating}.
The SF110 benchmark\footnote{\url{http://www.evosuite.org/experimental-data/sf110/}} 
is a set of Java classes, extracted from 100 projects in
the SourceForge repository \cite{rojas2017detailed, shamshiri2015random}.
We randomly sampled Java classes from the benchmarks discarding the trivial ones \cite{panichella2018automated}, \ie the classes having cyclomatic complexity below five.
In total, we selected 110 Java classes from 27 different projects, having 29,842 branches and 139,519 mutants considered as target coverage in our experiment.
Table \ref{table:subjects} reports the characteristics of the classes grouped by project.

\subsection{Experimental Protocol}
\label{sec:study:protocol}
We run \dmosa, \pdmosa, and random search for each class in the benchmark, collecting the resulting code coverage and mutation score. 
For this, the generated test cases/suite are post-processed in \evosuite: 
input data values and method sequences are minimized after the search process terminates. More precisely, redundant statements that do not satisfy any additional coverage targets (e.g., branches) are discarded. 
These post-processing steps are applied for both search algorithms under study~\cite{fraser2013whole}.
We set the maximum search time to 180 seconds \cite{campos2017empirical}. 
Hence, the search stops either if the 100\% coverage is reached or the time budget runs out.
We set an extra timeout of 10 minutes at the end of the search for mutation analysis.
We use this budget because of the additional overhead required to re-execute each test case against the target mutants.
To deal with the non-determinism of the employed algorithms, each run is repeated 50 times \cite{campos2017empirical}.
We adopt the default GA parameters used by \evosuite\cite{fraser2013whole} since they provide good results \cite{arcuri2013parameters}.

We rely on the non-parametric Wilcoxon Rank-Sum Test~\cite{conover1999practical} with significance level $\alpha$=0.05.
We formulate three null hypotheses, one for each research question, \ie that the compared algorithms achieve the same target coverage (\RQ{1}), the same strong mutation coverage (\RQ{2}), and the same runtime and heap memory consumption (\RQ{3}).
$p$-values$<$0.05 allow us to reject these null hypotheses.
Moreover, we rely on the Vargha-Delaney ($\hat{A}_{12}$) statistic~\cite{vargha2000acritique} to estimate the effect size of the differences between the achieved distributions.
It has the following interpretation: 
for the coverage criteria and mutation score $\hat{A}_{12}$ $\geq$ 0.50 when \dmosa ---or the random search--- achieves a higher coverage than \pdmosa while $\hat{A}_{12}$ $<$ 0.50 means the opposite.
For runtime and memory consumption $\hat{A}_{12} \geq 0.50$ indicates that the suites generated by \pdmosa run faster or use less memory than the ones generated by \dmosa.
Vargha-Delaney ($\hat{A}_{12}$) statistic also returns a categorical estimation of the effect size values~\cite{vargha2000acritique}, with \textit{negligible, small, medium}, and \textit{large} as possible levels.

\medskip
\noindent\textbf{Mutation-based analysis}.
To evaluate the fault detection effectiveness of \pdmosa, we rely on strong mutation analysis, due to several reasons:
(i) Multiple studies showed a significant correlation between mutant detection and real-fault detection~\cite{just2014mutants, andrews2005mutation}.
(ii) Mutation testing is broadly recognized as an upscale coverage criterion~\cite{jia2011analysis}, and it was shown to be a superior measure of test case effectiveness compared to other criteria~\cite{wei2012branch, inozemtseva2014coverage}.
The underlying idea of mutation testing is the creation of modified versions of the original source code, called \emph{mutants}~\cite{offutt2011mutation}.
These changes are introduced in the production code by \emph{mutation operators}, aiming to mime real faults \cite{just2014mutants}.
(iii) Each test suite is run against the generated mutants and evaluated based on its \emph{mutation score}, \ie the ratio of killed (detected) mutants and the number of generated ones.

To perform our analysis, we rely on \textsc{EvoSuite}'s built-in mutation engine~\cite{fraser2015achieving}, implementing eight different mutation operators, \ie \emph{Delete Call}, \emph{Delete Field}, \emph{Insert Unary Operator}, \emph{Replace Arithmetic Operator}, \emph{Replace Bitwise Operator}, \emph{Replace Comparison Operator}, \emph{Replace Constant}, and \emph{Replace Variable}.
We opt for \textsc{EvoSuite}'s engine for two reasons:
First, it makes strong mutation analysis straightforward.
Second, it was shown that the mutation scores computed by \evosuite are close to results on real-world software~\cite{fraser2015achieving}, which motivated recent works to rely on it~\cite{panichella2018automated, panichella2018incremental}.

\medskip
\noindent\textbf{Performance measurement}. 
To evaluate the performance, we compare the runtimes and heap memory usages of the test suites generated by \dmosa and \pdmosa.
An ideal comparison would require measuring two identical test suites ---for each subject and approach--- in terms of branch coverage and statements executed.
However, this is impossible in practice due to the algorithms' randomness.
To have a fair comparison, we conduct our performance analyses selecting the test suites with statistically equivalent branch coverage.
We first select the classes with no statistical difference (\ie $p$-value$>$0.05) in branch coverage;
and then for each subject and approach, we select the test suites with the \emph{median} coverage over 50 runs for performance profiling and comparison. The median coverage was preferred over the average, because using the average coverage could results in slightly more diverse (i.e., incomparable) test cases in terms of actual coverage.
To profile the performance of the test suites, we proceed as follows:
We transform the source code files for performance measurements, compile the augmented versions, and run the test suites with the \textsc{EvoSuite} standalone runtime.
The transformer employs JavaParser\footnote{\url{https://github.com/javaparser/javaparser}} 
for AST transformations.
It adds for every test case a method before (\texttt{@Before}) and after (\texttt{@After})
its execution, which reports the current performance counters.
These counters, as reported by Java's MXBeans 
(\texttt{RuntimeMXBean}, \texttt{MemoryMXBean}, \texttt{GarbageCollectorMXBean}, and \texttt{OperatingSystemMXBean}), are:
the current time stamp (in nanoseconds), 
the heap size (in bytes),
the \gls{gc} count (number of garbage collections since the virtual machine started),
and the \gls{gc} time (in milliseconds).
We executed the performance measurements on a bare-metal machine  
reserved exclusively for the measurements, \ie 
without user-level background processes (except \texttt{ssh}) running.
The machine has a 12-core Intel Xeon X5670@2.93GHz CPU with 70 GiB memory, 
runs ArchLinux with a kernel version 5.2.9-arch1-1-ARCH,
and uses a Samsung SSD 860 PRO SATA III disk.

We execute and measure each test suite 1000 times (forks), in a fresh \gls{jvm}, resembling the methodology proposed by Georges~\etal~\cite{georges2007}.
In a post-processing step,
we compute the diffs for each test case and calculate the sum of all test cases to retrieve the overall performance (\ie runtime and heap size) of each test suite.
As heap memory diffs might be influenced by \gls{gc} activity and therefore invalid, we replace the heap memory diff of affected methods
with the \emph{median} of the \emph{other} forks' valid results (\ie not affected by \gls{gc} activity).

\section{Results \& Discussion}
\label{sec:study:result}
This section discusses the results of the study answering the research questions formulated in \cref{sec:study}.
In the following, we will only refer to 109 classes since 1 class in our sample led to EvoSuite crashes caused by internal errors~\cite{panichella2018automated}.

\subsection{RQ1 - \rqoneshort}
\label{sec:rq1-results}
\begin{table*}[t!]
\centering
\caption{Comparison between Random Search, DynaMOSA, and aDynaMOSA on the considered criteria}
\label{table:criteria-comparison}
\resizebox{\linewidth}{!}{
\begin{tabular}{@{}lrrrrrrrrr@{}}
\toprule
         \textbf{Criterion} &  
         \multicolumn{3}{c}{\textbf{Average Coverage}} & 
         \multicolumn{3}{c}{\textbf{Random vs aDynaMOSA}} &
         \multicolumn{3}{c}{\textbf{DynaMOSA vs aDynaMOSA}} \\
         \cmidrule(r){2-4} \cmidrule(r){5-7} \cmidrule{8-10}
         & \textbf{Random} & \textbf{DynaMOSA} & \textbf{aDynaMOSA} & \textbf{\#Better} & \textbf{\#Worse} & \textbf{\#No Diff.} & \textbf{\#Better} & \textbf{\#Worse} & \textbf{\#No Diff.} \\
\midrule
 Branch        & 0.67 & 0.72 & \textbf{0.72} & 77  & 5 & 27 & 11 & 19 & 79 \\
 Line          & 0.71 & 0.76 & \textbf{0.77} & 80  & 2 & 27 & 16 & 17 & 76 \\
 Weak Mutation & 0.71 & 0.77 & \textbf{0.78} & 91  & 0 & 18 & 15 & 15 & 79 \\
 Method        & 0.93 & \textbf{0.97} & 0.97 & 59  & 0 & 50 & 2  & 4  & 103 \\
 Input         & 0.55 & \textbf{0.83} & 0.83 & 103 & 0 & 6  & 16 & 11 & 82 \\
 Output        & 0.41 & 0.54 & \textbf{0.54} & 91  & 0 & 18 & 22 & 7  & 80 \\
 Exception     & \textbf{0.99} & 0.98 & 0.99 & 3   & 7 & 99 & 14 & 3  & 92 \\
\bottomrule
\end{tabular}
}
\end{table*}
Table \ref{table:criteria-comparison} summarizes the code coverage achieved by random search, \dmosa, and \pdmosa according to the different coverage criteria. 
For each approach, we report 
(i) the mean coverage for each criterion over the 109 CUTs and
(ii) the number of classes for which \pdmosa is statistically better, worse, or equivalent than random search and \dmosa, according to the Wilcoxon test.
Furthermore, we discuss the $\hat{A}_{12}$ effect size.
Full results at the class level are reported in the replication package~\cite{appendix}.
\begin{figure}[t]
	\centering
	\includegraphics[width=\columnwidth]{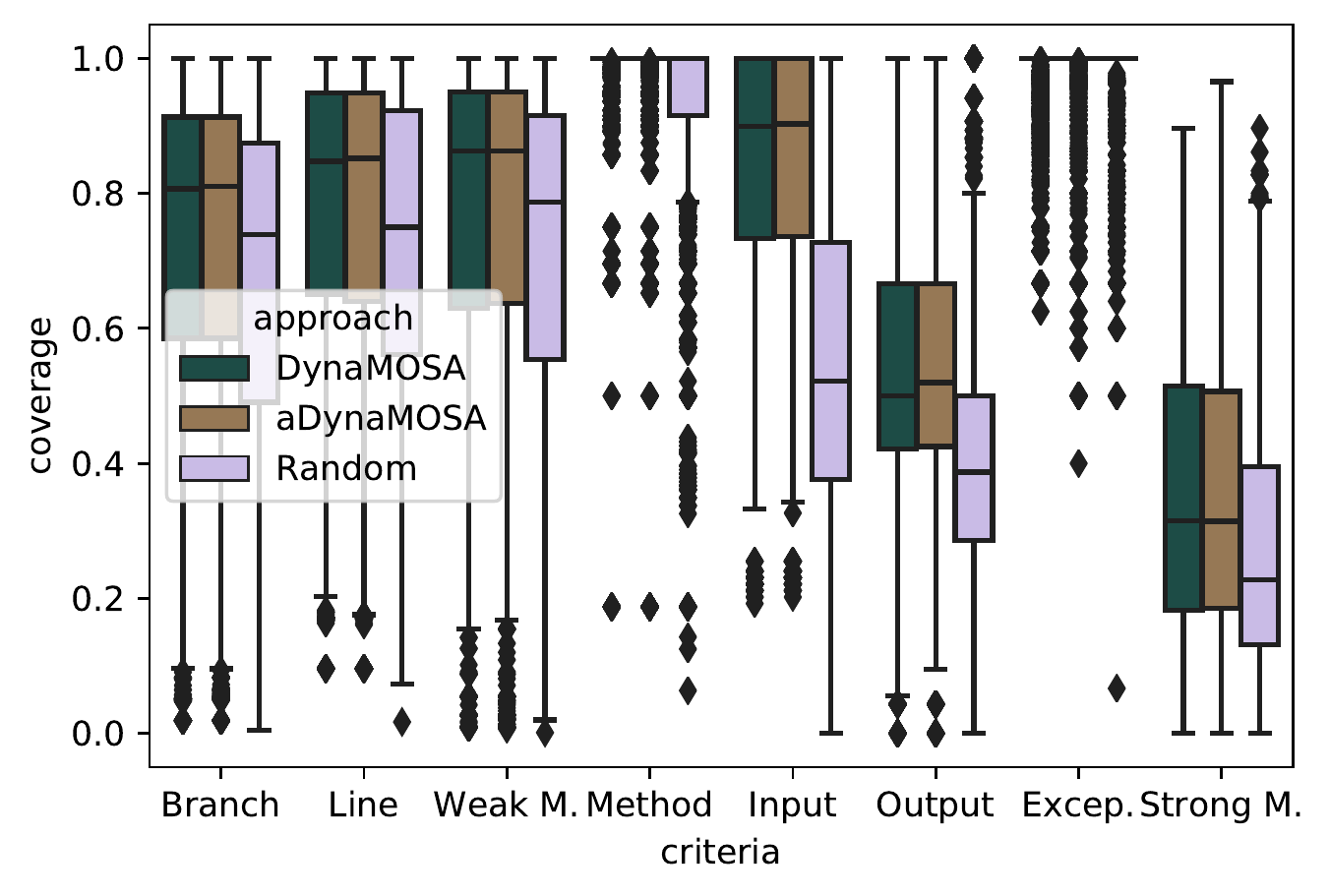}
	\caption{Comparison of target coverage achieved by Random Search, \dmosa, and \pdmosa over 50 independent runs for the 109 studied subjects.}
	\label{fig:coverage}
\end{figure}

\smallbreak
Table~\ref{table:criteria-comparison} compares \dmosa and \pdmosa.
For branch coverage, the two algorithms are almost equivalent:
on average, they achieve the same coverage (\ie \textasciitilde72\%) with a median of about 81\% and 80\%, respectively for \pdmosa and \dmosa.
The latter is statistically significantly better for 19 out of 109 classes (\textasciitilde18\%);
over such classes, it achieves +2 percentage points (\emph{pp}) in term of branch coverages.
Vice versa, \pdmosa significantly outperforms \dmosa in 11 out of 109 classes (\textasciitilde10\%) with an average difference of +3\emph{pp}.
For the vast majority of the subjects, (\ie \textasciitilde~78\%) there is no statistical difference between the two algorithms.
Similar results can be observed for line and weak mutation coverage, where for the 70\% and 72\% of the subjects the two approaches do not show a statistically significant difference, respectively.
\pdmosa only achieves on average +1\emph{pp} for both line and weak mutation coverage over the entire set of classes: \ie \textasciitilde77\% vs. \textasciitilde76\% and \textasciitilde78\% vs. \textasciitilde77\%, respectively for line and weak mutation coverage.
For the remaining criteria, \ie method, input, output, and exception, the number of subjects with no statistically significant difference increases,
ranging from 73\% to 94\% of the CUTs.
For only three classes, \dmosa covers more exceptions than \pdmosa.
Overall, for none of the investigated coverage criteria we observe large differences between \dmosa and \pdmosa.

\smallbreak
Comparing branch coverage of \pdmosa and random search (Table~\ref{table:criteria-comparison}), \pdmosa achieves on average +5\emph{pp} over all the subjects.
77 out of 109 classes are statistically significantly better (\textasciitilde70\%), while only 5 out of 109 classes are worse.
For 62 of these 77 subjects, the magnitude of the difference is \emph{large}.
The largest improvement is obtained for \texttt{ICSSearchAlgorith} (\texttt{weka} project) where \pdmosa covers 31\% more branches on average.
We observe similar results for all other criteria but the exception coverage where random search is not statistically significantly different for 103 out of 109 subjects.
\pdmosa achieves +6\emph{pp} and +7\emph{pp} for line and weak mutation coverage, respectively;
while \pdmosa reaches +4\emph{pp} for the method coverage criterion, it achieves +28\emph{pp} and +13\emph{pp} for input and output coverage, respectively.

\smallbreak
Figure \ref{fig:coverage} depicts an overview of the coverage scores achieved by the three approaches over the distinct criteria.
It highlights that \dmosa and \pdmosa have similar distributions for the different target coverages.
On the other and, except for the exception coverage, \pdmosa leads to larger coverage scores compared to the random search.

\resultsummary{\textbf{Finding 1.} \textit{Across seven criteria, \pdmosa achieves similar levels of coverage compared to \dmosa, while both outperform random search.}}


\subsection{RQ2 - \rqtwoshort}

    \begin{table*}[t]
    \centering
    \caption{Mean mutation score achieved for each project}
    \label{table:mutation-score}
    \resizebox{\linewidth}{!}{
    \begin{tabular}{@{}lr rrr rrr rrr@{}}
    \toprule    

    \textbf{Project} &
    \textbf{Classes} &
    \multicolumn{3}{c}{\textbf{Mutation Score}} &
    \multicolumn{6}{c}{\textbf{$\hat{A}_{12}$ Statistics}} \\
    \cmidrule(r){3-5} \cmidrule{6-11}

    ~ &
    ~ &   
    \textbf{Random} &
    \textbf{DynaMOSA} &
    \textbf{aDynaMOSA} &
    \multicolumn{3}{c}{\textbf{aDynaMOSA vs Random}} &
    \multicolumn{3}{c}{\textbf{aDynaMOSA vs DynaMOSA}} \\ 
    \cmidrule(r){6-8} \cmidrule{9-11}
    ~ & ~ & ~ & ~ & ~ & \textbf{\#Better} & \textbf{\#Worse} & \textbf{\#No Diff.} & \textbf{\#Better} & \textbf{\#Worse} & \textbf{\#No Diff.} \\ 

    \midrule
 freehep         & 6  & 0.25 & \textbf{0.37} & 0.36          & 5  & 0 & 1 & 1 & 2 & 3 \\
 fastjson        & 10 & 0.28 & \textbf{0.36} & 0.35          & 10 & 0 & 0 & 1 & 2 & 7 \\
 weka            & 10 & 0.21 & 0.24          & \textbf{0.25} & 9  & 0 & 1 & 2 & 1 & 7 \\
 re2j            & 7  & 0.31 & 0.34          & \textbf{0.35} & 5  & 0 & 2 & 1 & 1 & 5 \\
 bcel            & 4  & 0.33 & \textbf{0.42} & 0.39          & 3  & 0 & 1 & 1 & 2 & 1 \\
 wheelwebtool    & 7  & 0.24 & \textbf{0.33} & 0.32          & 7  & 0 & 0 & 2 & 3 & 2 \\
 javathena       & 1  & 0.22 & 0.24          & \textbf{0.25} & 1  & 0 & 0 & 1 & 0 & 0 \\
 math            & 4  & 0.31 & 0.35          & \textbf{0.36} & 4  & 0 & 0 & 2 & 0 & 2 \\
 image           & 7  & 0.29 & \textbf{0.37} & 0.36          & 6  & 0 & 1 & 2 & 1 & 4 \\
 webmagic        & 4  & 0.40 & 0.43          & \textbf{0.44} & 3  & 0 & 1 & 0 & 0 & 4 \\
 jdbacl          & 2  & 0.37 & \textbf{0.44} & 0.43          & 2  & 0 & 0 & 0 & 1 & 1 \\
 okio            & 9  & 0.24 & 0.32          & \textbf{0.34} & 8  & 0 & 1 & 2 & 0 & 7 \\
 okhttp          & 5  & 0.28 & \textbf{0.34} & 0.34          & 5  & 0 & 0 & 0 & 1 & 4 \\
 shop            & 4  & 0.34 & \textbf{0.40} & 0.39          & 4  & 0 & 0 & 0 & 3 & 1 \\
 jsecurity       & 1  & 0.29 & 0.34          & \textbf{0.34} & 1  & 0 & 0 & 0 & 0 & 1 \\
 fixsuite        & 1  & 0.06 & 0.09          & \textbf{0.12} & 1  & 0 & 0 & 0 & 0 & 1 \\
 javaviewcontrol & 2  & 0.12 & \textbf{0.16} & 0.15          & 2  & 0 & 0 & 0 & 1 & 1 \\
 byuic           & 1  & 0.08 & \textbf{0.11} & 0.10          & 1  & 0 & 0 & 0 & 1 & 0 \\
 gson            & 4  & 0.16 & \textbf{0.21} & 0.19          & 3  & 0 & 1 & 0 & 2 & 2 \\
 firebird        & 3  & 0.48 & \textbf{0.52} & 0.51          & 3  & 0 & 0 & 0 & 1 & 2 \\
 jxpath          & 3  & 0.51 & 0.55          & \textbf{0.57} & 3  & 0 & 0 & 1 & 0 & 2 \\
 a4j             & 2  & 0.24 & 0.19          & \textbf{0.27} & 1  & 0 & 1 & 2 & 0 & 0 \\
 jmca            & 2  & 0.19 & \textbf{0.29} & 0.28          & 2  & 0 & 0 & 0 & 0 & 2 \\
 freemind        & 1  & 0.19 & 0.21          & \textbf{0.25} & 1  & 0 & 0 & 1 & 0 & 0 \\
 la4j            & 7  & 0.25 & 0.33          & \textbf{0.34} & 3  & 0 & 4 & 0 & 0 & 7 \\
 saxpath         & 1  & 0.57 & \textbf{0.60} & 0.59          & 1  & 0 & 0 & 0 & 1 & 0 \\
 jiprof          & 1  & 0.06 & 0.13          & \textbf{0.13} & 1  & 0 & 0 & 0 & 0 & 1 \\
\midrule
\multicolumn{2}{@{}l@{}}{Mean over projects} & 0.27 & 0.32 & 0.33 & \multicolumn{6}{c}{~} \\\midrule
\multicolumn{5}{@{}l@{}}{No. cases aDynaMOSA is better than Random} & \multicolumn{4}{c}{~} & \multicolumn{2}{@{}r@{}}{95 (87.16\%)} \\ 
\multicolumn{5}{@{}l@{}}{No. cases Random is better than aDynaMOSA} & \multicolumn{4}{c}{~} & \multicolumn{2}{@{}r@{}}{0 (0.0\%)} \\ 
\multicolumn{5}{@{}l@{}}{No. cases aDynaMOSA is better than DynaMOSA} & \multicolumn{4}{c}{~} & \multicolumn{2}{@{}r@{}}{19 (17.43\%)} \\  
\multicolumn{5}{@{}l@{}}{No. cases DynaMOSA is better than aDynaMOSA} & \multicolumn{4}{c}{~} & \multicolumn{2}{@{}r@{}}{23 (21.1\%)} \\ 
\bottomrule
\end{tabular}
}
\end{table*}
Figure~\ref{fig:coverage} shows the distributions of the mutation scores (\ie strong mutation coverage) using box-plots ---on the extreme right, along with the other criteria--- achieved by the approaches for the 109 subjects over 50 runs.
We notice that the distributions of \pdmosa and \dmosa are similar:
the former achieves +1\emph{pp} mutation scores compared to the baseline.
The medians of the distributions are \textasciitilde33\% and \textasciitilde32\%, respectively. 
Both approaches considerably outperform random search which achieves a median of \textasciitilde27\%.

Table~\ref{table:mutation-score} reports the fine-grained results of strong mutation achieved by random search, \dmosa, and \pdmosa.
We report 
(i) the mutation scores averaged over the different projects and
(ii) the number of cases in each project where \pdmosa is better, worse, or equivalent ---according to the Wilcoxon test--- compared to random search and \dmosa.
We share the full results at class level in the replication package~\cite{appendix}.

\smallbreak
From Table~\ref{table:mutation-score} we observe that \pdmosa significantly outperforms random search in 95 out of 109 cases, corresponding to \textasciitilde87\% of all the CUTs.
For these subjects, the test suites generated by \pdmosa achieve from +1\emph{pp} to +29\emph{pp} higher mutation scores compared to the ones generated by the baseline, with an average improvement of \textasciitilde7\%.
In 78 out of these 95 cases, the magnitude of the difference is \emph{large}.
Random search is never significantly better than \pdmosa.

\smallbreak
Comparing \pdmosa and \dmosa, in more than half of the cases (\ie \textasciitilde61\%) there is no statistical difference between the mutation score of the two approaches, which is in line with what we observed in RQ1.
For \textasciitilde23\% of the subjects, \dmosa scores a significantly higher in strong mutation.
However, for about half of these cases (10 out of 23 subjects) the magnitude of the difference is \emph{small}.
The suites generated by \dmosa achieve from +0.3\emph{pp} to +8\emph{pp} higher mutation score, with an average improvement of \textasciitilde3\%.
On the other hand, in \textasciitilde17\% of the subjects, \pdmosa outperforms the baseline.
In these, 19 out of 109, the suites generated by \pdmosa achieve from +0.8\emph{pp} to +15\emph{pp} (for the class \texttt{Product}) higher mutation score, with a mean of \textasciitilde4\%. 

In the few cases where \pdmosa performs worse than \dmosa, the difference is due to a slight difference in branch coverage.
There is a direct relation between code coverage and fault effectiveness:
if a mutant is not covered, it cannot be killed.
For example, let us consider the class \texttt{Parser} from \texttt{re2j}, which has 667 branches and 501 mutants. \pdmosa achieves 63.0\% average branch coverage compared to 63.7\% achieved by \dmosa.
However, neither set of mutants killed by one of the two approaches is a subset of the other approach's set.
\pdmosa kills nine mutants not killed by \dmosa while \dmosa kills 18 mutants not killed by \pdmosa. 
Listing~\ref{code:example} shows an example of mutants killed by \dmosa only.
\begin{lstlisting}[caption={The listing shows a mutant generated by EvoSuite's mutation engine. While the suites generated by \dmosa are able to it, the mutant survives to the ones generated by \pdmosa.}, label={code:example},float={t!}]
public class Parser {
	...
	private Regexp removeLeadingString(Regexp re, int n) {
  	// original code:
  	// if ((re.op == Regexp.Op.CONCAT) && (re.subs.length == 0))
  	// mutant:
  	if ((op == Regexp.Op.CONCAT) && (subs.length > 0)) { ... }
   	...
}
\end{lstlisting}

The mutant is injected into the first \texttt{if} statement of the private method \texttt{removeLeadingString}.
Both approaches cover the statement through indirect method calls; however, only the test cases produced with \dmosa are able to kill the mutant.
The reason for this is that the if-condition requires to instantiate an object of class \texttt{Regexp} with proper attributes \texttt{op} and \texttt{subs}.
This can be done by invoking additional methods of \texttt{Regexp}.
\pdmosa is designed to reduce the number of method calls (to reduce heap memory consumption); therefore, in some runs, it generates tests without setting the input object \texttt{re}.
This example suggests that there is room for further improvement of \pdmosa by handling method calls differently, depending on whether they are required for fixture or for exercising the CUT behavior.

\resultsummary{\textbf{Finding 2.} \textit{\pdmosa achieves similar levels of mutation score compared to \dmosa, while both outperform random search.}}3

\subsection{RQ3 - \rqthreeshort}

    \begin{table*}[t]
    \centering
    \caption{Mean runtime and memory consumption achieved for each project}
    \label{table:performance}
    \resizebox{\linewidth}{!}{
    \begin{tabular}{@{}lr rr rr rrr rrr@{}}
    \toprule    

    \textbf{Project} &
    \textbf{Classes} &
    \multicolumn{2}{c}{\textbf{Runtime (in ms)}} &
    \multicolumn{2}{c}{\textbf{Memory Consumption (in)}} &
    \multicolumn{6}{c}{\textbf{$\hat{A}_{12}$ Statistics}} \\
    \cmidrule(r){3-4} \cmidrule(r){5-6} \cmidrule{7-12}

    ~ &
    ~ &   
	\textbf{DynaMOSA} &
    \textbf{aDynaMOSA} & 
    \textbf{DynaMOSA} &
    \textbf{aDynaMOSA} &
    \multicolumn{3}{c}{\textbf{Runtime}} &
    \multicolumn{3}{c}{\textbf{Memory Consumption}} \\ 
    \cmidrule(r){7-9} \cmidrule{10-12}
    ~ & ~ & ~ & ~ & ~ & ~ & \textbf{\#Better} & \textbf{\#Worse} & \textbf{\#No Diff.} & \textbf{\#Better} & \textbf{\#Worse} & \textbf{\#No Diff.} \\ 

    \midrule
 jmca            & 2 & \textbf{130.87} & 155.12          & \textbf{590.46} & 607.21          & 1 & 1 & 0 & 1 & 1 & 0 \\
 jdbacl          & 1 & 14.32           & \textbf{2.28}   & \textbf{2.84}   & 5.68            & 0 & 0 & 1 & 0 & 0 & 1 \\
 javaviewcontrol & 1 & 169.81          & \textbf{112.79} & 244.47          & \textbf{204.48} & 1 & 0 & 0 & 1 & 0 & 0 \\
 jsecurity       & 1 & 552.76          & \textbf{490.33} & 392.03          & \textbf{349.12} & 1 & 0 & 0 & 1 & 0 & 0 \\
 freemind        & 1 & \textbf{781.37} & 1,188.46        & \textbf{305.05} & 384.81          & 0 & 1 & 0 & 0 & 1 & 0 \\
 shop            & 3 & \textbf{57.18}  & 75.42           & 285.34          & \textbf{281.42} & 2 & 1 & 0 & 2 & 1 & 0 \\
 bcel            & 1 & 91.87           & \textbf{84.41}  & 568.01          & \textbf{488.62} & 1 & 0 & 0 & 1 & 0 & 0 \\
 a4j             & 1 & 49.26           & \textbf{48.75}  & \textbf{22.78}  & 22.83           & 1 & 0 & 0 & 0 & 0 & 1 \\
 firebird        & 2 & 83.36           & \textbf{50.27}  & 270.44          & \textbf{235.88} & 2 & 0 & 0 & 2 & 0 & 0 \\
 fastjson        & 7 & 444.47          & \textbf{174.20} & 1,129.95        & \textbf{515.34} & 6 & 1 & 0 & 4 & 3 & 0 \\
 webmagic        & 2 & \textbf{121.20} & 172.61          & 281.55          & \textbf{258.76} & 1 & 1 & 0 & 1 & 1 & 0 \\
 okio            & 7 & 125.32          & \textbf{105.05} & 524.28          & \textbf{456.86} & 4 & 2 & 1 & 7 & 0 & 0 \\
 math            & 2 & 295.59          & \textbf{286.86} & \textbf{181.71} & 177.68          & 1 & 1 & 0 & 1 & 1 & 0 \\
 image           & 5 & 53.51           & \textbf{50.62}  & 290.56          & \textbf{261.89} & 3 & 2 & 0 & 4 & 1 & 0 \\
 jxpath          & 2 & 126.13          & \textbf{66.42}  & 196.70          & \textbf{191.80} & 2 & 0 & 0 & 1 & 1 & 0 \\
 gson            & 2 & 72.53           & \textbf{45.51}  & 199.58          & \textbf{178.30} & 2 & 0 & 0 & 2 & 0 & 0 \\
 freehep         & 4 & 229.75          & \textbf{172.33} & 152.33          & \textbf{136.53} & 4 & 0 & 0 & 3 & 1 & 0 \\
 la4j            & 6 & 215.93          & \textbf{164.51} & 238.67          & \textbf{226.45} & 6 & 0 & 0 & 4 & 1 & 1 \\
 re2j            & 4 & 50.12           & \textbf{48.81}  & 262.36          & \textbf{254.76} & 2 & 2 & 0 & 3 & 1 & 0 \\
 okhttp          & 2 & 49.18           & \textbf{45.46}  & 129.87          & \textbf{125.13} & 1 & 1 & 0 & 2 & 0 & 0 \\
 weka            & 1 & \textbf{0.49}   & 389.94          & \textbf{0.00}   & 92.30           & 0 & 0 & 1 & 0 & 1 & 0 \\
\midrule
\multicolumn{2}{@{}l@{}}{Mean over projects} & 176.91 & 187.15 & 298.52 & 259.8             & 41 (71.93\%)   & 13 (22.81\%)   & 3 (5.26\%)             & 40 (70.18\%) & 14 (24.56\%) & 3 (5.26\%) \\\bottomrule
\end{tabular}
}
\end{table*}
In this section, we compare the runtime and heap memory consumption of the test suites generated by \pdmosa and \dmosa.
Recall that we restrict this analysis to the CUTs with no statistical difference in branch coverage (see~\cref{sec:study:protocol});
thus, we pick the suite with the median coverage for each subject.

%

Table~\ref{table:performance} summarizes the performance results of the suites generated by the two approaches, aggregated by project.
We first discuss the cases where \pdmosa outperforms \dmosa.
Table~\ref{table:performance} shows that the test suites generated by \pdmosa have a shorter runtimes in about 72\% of the cases.
For these suites, runtime decreases on average by \textasciitilde24\% (with a median of \textasciitilde13\%), ranging from -1\emph{pp} to -79\emph{pp} (for the class \texttt{JSONArray}).
Regarding heap memory consumption, \pdmosa outperforms \dmosa for \textasciitilde70\% of the classes.
Among these subjects, the suites generated by \pdmosa show a \textasciitilde15\% decrease in heap memory consumption (with a median of \textasciitilde11\%), ranging from -1.6\emph{pp} to -86\emph{pp} (for the class \texttt{JSONArray}).

\begin{figure}[t]
	\centering
	\includegraphics[width=\columnwidth]{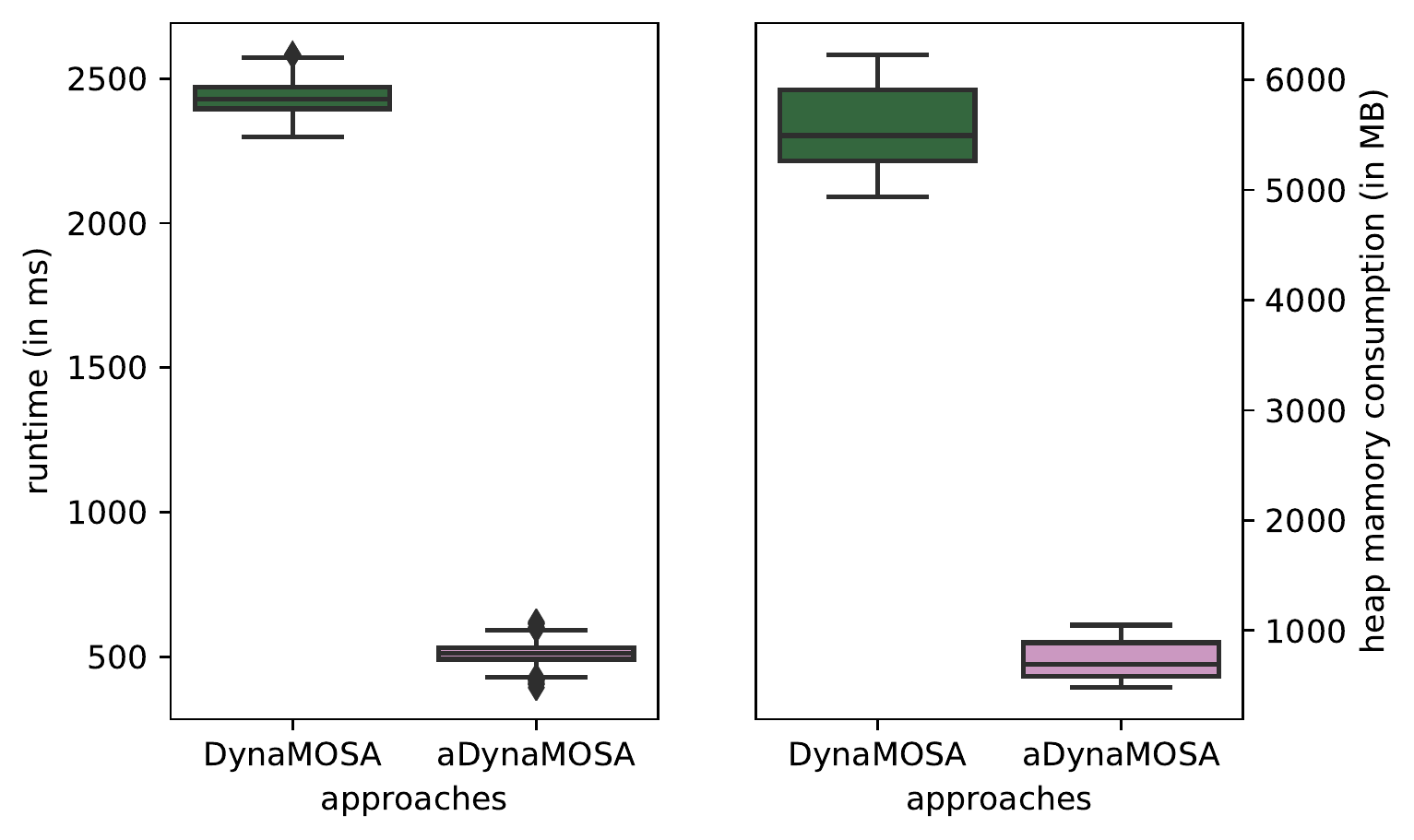}
	\caption{Comparison of runtime and heap memory consumption for the suite generated by \dmosa and \pdmosa for the \texttt{JSONArray} class over 1,000 independent runs.}
	\label{fig:performance-example}
\end{figure}

Figure~\ref{fig:performance-example} shows the example of the class \texttt{JSONArray}, plotting the runtime and heap memory consumption distributions over 1000 independent runs of the generated suites by \dmosa and \pdmosa.
The two profiled suites achieve similar levels of coverage over the seven different criteria.
However, the median runtime is \textasciitilde511~milliseconds~(ms) for \pdmosa versus \textasciitilde2,429~ms for \dmosa, while the median heap memory consumption is \textasciitilde694 megabyte (MB) for \pdmosa versus \textasciitilde5,429~MB for \dmosa.

To have a fair performance analysis, we compare test suites achieving the closest ---ideally identical--- branch coverage (see~\cref{sec:study:protocol}).
However, due to intrinsic randomnesses of GAs, this is \emph{practically infeasible}.
Thus, we extend our analysis by looking at the differences in code coverage for the cases in which the suites generated by \pdmosa show lower resource demands than the ones generated by \dmosa.
The goal is to verify whether the decreased runtime (or heap memory consumption) is caused by tests having lower code coverage, which may result in fewer statements being executed.
Despite faster runtime and lower memory consumption, this analysis shows \emph{there is no decrease on any target criterion}.
This is evident if we look at the median and the mean of differences in branch and line coverage achieved by the two approaches.
The mean of the differences is below 1\% ---0.5\% and 0.8\%, respectively for branch and line coverage--- while the median of the differences is exactly 0.0\% for both the criteria.

\resultsummary{
\textbf{Finding 3.} 
	\textit{The suites generated by \pdmosa have lower runtimes and heap memory consumption, respectively, for \textasciitilde72\% and \textasciitilde70\% of the subjects while achieving the same levels of code coverage.}
}

Regarding the CUTs where the test suites generated by \dmosa have better performance than the ones generated by \pdmosa:
Table~\ref{table:performance} shows that the suites generated by \dmosa outperform the ones generated by \pdmosa in \textasciitilde22\% of the (negative) cases.
For these subjects, the runtime of \dmosa decreases on average by \textasciitilde20\%, ranging from -0.5\emph{pp} to -48\emph{pp} (for the class \texttt{JSTerm}).
Similarly, the test suites generated by \dmosa show lower heap memory consumption for \textasciitilde24\% of the cases, with a decrease ranging from -0.2\emph{pp} to -39\emph{pp} (for the class \texttt{JSONLexerBase}).

\begin{figure}[t]
	\centering
	\includegraphics[width=\columnwidth]{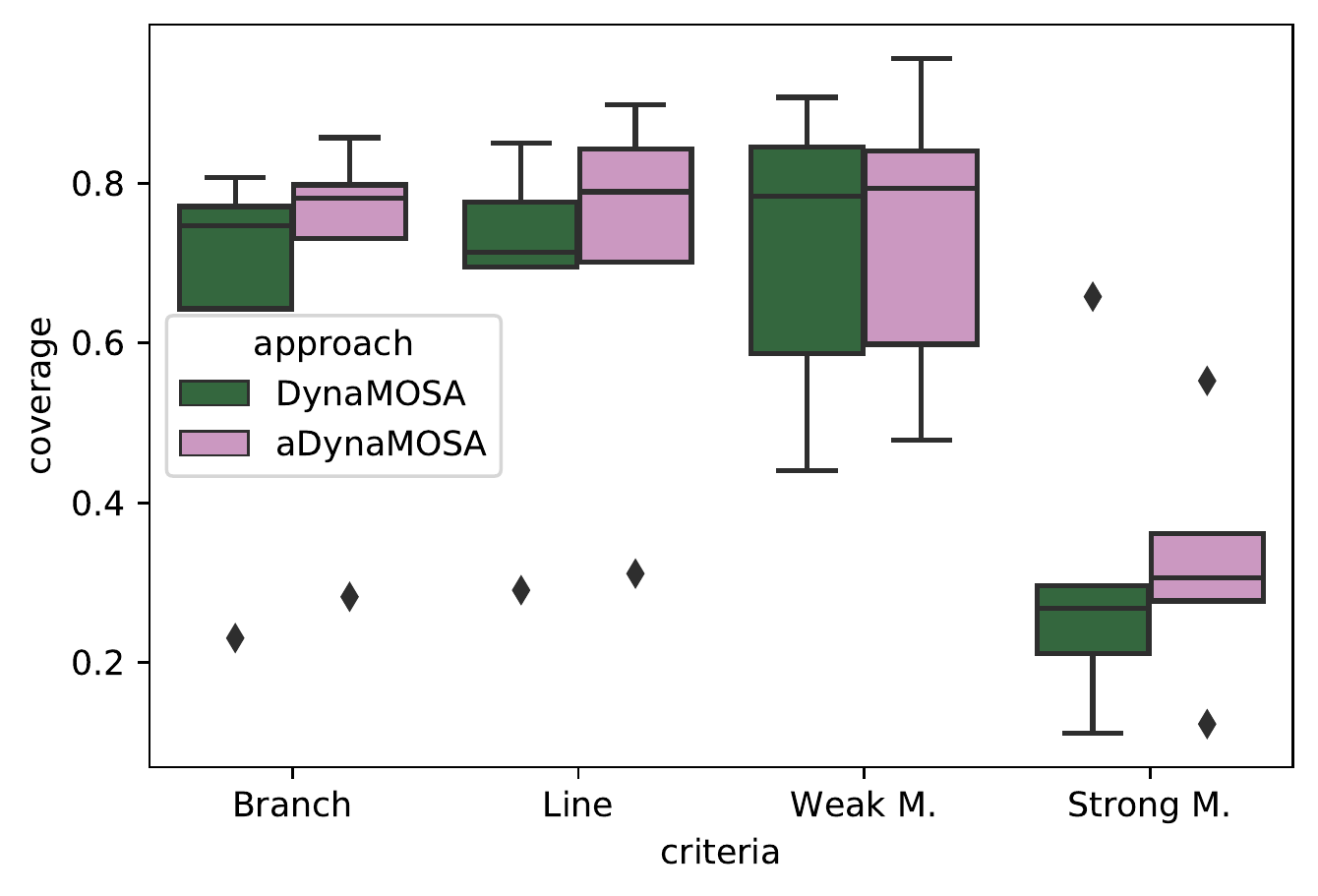}
	\caption{Coverage scores achieved by the test suites generated by the two approaches for the CUTs where \dmosa shows better performance than \pdmosa.}
	\label{fig:coverages-bad-cases}
\end{figure}

To understand the reason of these few negative results, we analyze the code coverage achieved by both approaches. 
Figure~\ref{fig:coverages-bad-cases} shows the branch and line coverage and weak and strong mutation score distributions achieved for the subjects where \dmosa produces more performant tests.
%
%
Although \pdmosa generates suites with inferior performance it reaches higher code coverage, i.e., +4\emph{pp} branch and +9\emph{pp} line coverage.
Each suite generated by \pdmosa contains 17 tests more on average than the ones generated by the baseline (154 vs. 137).
This indicates that the suites generated by aDynaMOSA \emph{execute more production code}, thus, executing more statements, resulting in higher resource demands.

Recall that we profile the test suites with the median branch coverage for the subjects that do not show statistically significant differences (\ie $p$-value $>$ 0.05) (see~\cref{sec:study:protocol}).
This allows for fair profiling and comparison of test suites with similar code coverage.
However, with a significance level of $0.05$, we include subjects in the performance analysis that show marginal statistical significance (\ie 0.05 $< p$-value $< 0.1$) in branch coverage.
Consequently in these cases, we select and profile test suites with a non-negligible difference in code coverage.
In our analysis, we observe a marginal statistical difference in branch coverage for \textasciitilde60\% of the CUTs where the baseline outperforms \pdmosa in resource usage.
A representative example is the class \texttt{JSONLexerBase}, which has the worst heap memory consumption achieved by \pdmosa compared to the baseline.
However, this generated suite achieves +5\emph{pp} branch coverage and contains 34 more test cases on average.
\resultsummary{
\textbf{Finding 4.} 
\textit{
The suites generated by \pdmosa have slower runtimes and higher heap memory consumption for \textasciitilde22\% and \textasciitilde24\% of the subjects, respectively.
However, in these cases they achieve higher code coverage and contain more test cases.
}
}

\subsection{Discussion}
\label{sec:discussion}
%
%
Listing~\ref{code:mosa} depicts two test cases for the class \texttt{GaussianSolver} (from \texttt{la4j}) generated by DynaMOSA and aDynaMOSA.
The test case generated by the former is the one with both slower runtime and higher heap memory consumption having about 39ms and 38MB on average over 1000 runs.
First, it creates a \texttt{Matrix} with a diagonal and size equal to 1349 (line 3 of Listing~\ref{code:mosa}).
Second, it creates an object of the class \texttt{SparseVector} with size and capacity equal to 1349 (line 4 of Listing~\ref{code:mosa}). 
Thus, it instantiates an object of the class \texttt{GaussianSolver} from the  matrix above (line 5 of Listing~\ref{code:mosa}).
Finally, it executes the method \texttt{solve} that solves the corresponding linear system (line 6 of Listing~\ref{code:mosa}).
The test generated by aDynaMOSA for the same class builds the \texttt{GaussianSolver} using a very small matrix (line 13 of Listing~\ref{code:mosa}).
Similarly, a smaller \texttt{SparseVector} is then instantiated in line 14 of Listing~\ref{code:mosa}.
At the end, the \texttt{solve} method is again called to solve the linear system (line 16 of Listing~\ref{code:mosa}).

Despite implementing a similar behavior, the test generated by DynaMOSA runs almost 20 times slower (and using 8 times more heap memory) ---on average over the 1000 runs--- than the one generated by aDynaMOSA.
This improvement is due to a better input value selection for the methods directly or indirectly invoked by the generated tests.
While the algorithm has no direct control over this selection, the selective pressure applied by the performance proxies favors the individuals with better inputs ---from a performance perspective--- that randomly appear in the population.
For this reason, we expect aDynaMOSA to be particularly more effective in scenarios where the input space is not trivial (\ie most inputs are primitive values and the CUT does not handle large arrays or objects).

\begin{lstlisting}[caption={Test cases for the \texttt{GaussianSolver} class.}, captionpos=b, label={code:mosa},float={t}]
// test case generated by DynaMOSA
public void test03()  throws Throwable  {
  Matrix matrix = Matrix.unit(1349, 1349);
  SparseVector sparseVector = SparseVector.zero(1349);
  GaussianSolver gaussianSolver = new GaussianSolver(matrix);
  gaussianSolver0.solve(sparseVector);
}

// test case generated by aDynaMOSA
public void test12()  throws Throwable  {
  MockRandom mockRandom = new MockRandom(1);
  DenseMatrix denseMatrix = DenseMatrix.randomSymmetric(1, mockRandom);
  GaussianSolver gaussianSolver = new GaussianSolver(denseMatrix);
  SparseVector sparseVector = SparseVector.zero(1, 1);
  try { 
    gaussianSolver0.solve(sparseVector);
    ...
}
\end{lstlisting}

\smallbreak
\noindent\textbf{Need for an adaptive approach.}
As explained in \cref{sec:algo}, we considered an adaptive approach that disables/enables the performance heuristics depending on whether the search stagnates, \ie there is no improvement in the objective values for subsequent generations.
To provide empirical evidence for the need for an adaptive approach, we conducted an additional study by running \pdmosa and disabling the GET-SECONDARY-HEURISTIC procedure (see~\cref{sec:algo}):
\ie the algorithm always uses the heuristic based on the performance proxies.
Our results show the expected decrease in code coverage:
for branch coverage, the non-adaptive version of aDynaMOSA achieves on average -18\emph{pp} in 52 out of 109 cases (\textasciitilde48\%).
On the contrary, DynaMOSA never achieves higher branch coverage.
We observe a similar situation for weak mutation coverage:
the non-adaptive version of aDynaMOSA achieves on average -25\emph{pp} in 48 out of 109 cases (\textasciitilde44\%), while the opposite never happens.

\smallbreak
\noindent\textbf{Oracle cost.}
\emph{Quantitative human oracle cost reduction} aims at reducing test suite and test case size to consequently diminish the amount of human effort required to check the candidate assertions (the \emph{oracle problem})~\cite{barr2015oracle}.
Test suite size has often been used in literature as a proxy for oracle cost~\cite{harman2010optimizing, ferrer2012evolutionary}.
A simple solution for alleviating this issue is to reduce the size of the generated test suites~\cite{harman2010optimizing}.
To investigate the oracle cost for \dmosa and \pdmosa, we compare the size of the generated suite with the Wilcoxon test~\cite{conover1999practical}.
We observe that the test suites generated by aDynaMOSA are significantly smaller in 69 out of 109 cases (\textasciitilde63\%), while the opposite happens in only 6 cases.
The average test case length is 98 and 91 statements for DynaMOSA and aDynaMOSA, respectively.
These results give us confidence that our approach ---as a collateral effect--- might help to reduce the human oracle cost to a greater extent than DynaMOSA.
Note that the test suites generated by both \dmosa and \pdmosa are post-processed for test minimization. 
Therefore, the differences observed in terms of test suite size are due to the adaptive strategies and the performance proxies implemented in \pdmosa.
We report the full results at class level in the replication package~\cite{appendix}.

\smallbreak
\noindent\textbf{Trade-off between coverage and performance}. 
Our results show that aDynaMOSA achieves similar levels of coverage while optimizing runtime and memory consumption.
Despite aDynaMOSA finding a good compromise between primary and secondary objectives, in a few cases the performance optimization results in slightly lower coverage.  
The acceptable level of performance and coverage depends on the system domain. 
For instance, in the development context of cyber-physical systems (CPS), tests can be particularly expensive to run, especially when they involve hardware or simulations~\cite{torngren2018complexity}.
Thus, the resource demands for testing systems in this domain are dramatically higher compared to non-CPS-based applications.
Adaptive approaches focusing on performance while keeping high levels of coverage might improve the testability of CPS~\cite{asadollah2015survey, torngren2018complexity}.

\section{Threats to Validity}
\label{sec:threats}

Threats to \emph{construct validity} regard the way the performance of a testing technique is defined.
To compare the effectiveness of the different algorithms, we rely on metrics extensively exploited in the literature~\cite{panichella2018automated}.
For RQ1, we evaluate \pdmosa relying on the seven default criteria of EvoSuite, \ie branch, line, weak mutation, method, input, output, and exception coverage.
In RQ2, we use strong mutation coverage.
To give a reasonable estimation of the performance of the generated test suites, we use runtime and heap memory consumption in RQ3. 
The usage of different tools might influence the results.
To tackle this threat, all the algorithms we compare are implemented in EvoSuite~\cite{fraser2011evosuite}.

Threats to \emph{internal validity} concern lurking variables that might influence our results.
A common threat that arises dealing with genetic algorithms is related to their intrinsic randomness.
To deal with it, we repeated each run 50 times~\cite{campos2017empirical}.
We discuss the average results paired with statistical significance tests.
Different factors might have also influenced the performance measurements, such as  the order in which the tests are executed.
Due to dynamic compiler optimizations, different execution orders might change the runtime results of individual runs.
We tackle this threat by repeating the measurements for 1000 times.
Another threat concerns the memory measurements where garbage collector activity invalidates the heap diff computed for every test method.
We address this threat by replacing the measurements for the methods that trigger the GC with the \emph{other valid} forks' average heap utilization.
To lower the resources demand of generated tests, we aggregate seven different proxies in a performance score optimized as a secondary objective.
To investigate their impact in isolation, we run aDynaMOSA with a single proxy enabled at a time.
Then, we measure the runtime and the achieved branch coverage of the generated tests, averaged over five different runs (measured in EvoSuite).
While the average runtime varies across the different proxies, we observe that their usage in isolation always results in lower values of branch coverage compared to their usage in aggregation.

To investigate the oracle cost of aDynaMOSA, we compare the size of the generated suites to the ones produced by DynaMOSA.
Previous research relied on test suite size as a proxy for the oracle cost~\cite{harman2010optimizing}.
We can only empirically claim that aDynaMOSA generates smaller suites and are therefore confident that aDynaMOSA might help to reduce the oracle cost.
However, too many other factors prevent us from making a definitive claim.
Conducting a study with developers to confirm this assumption is part of our future agenda.

Threats to \emph{External Validity} regard the generalizability of the results.
We conduct our experiment on randomly selecting Java classes from four different datasets~\cite{fraser2014large, panichella2017java, panichella2018automated, panichella15reformulating} used in several previous works on test case generation.
In total, we selected 110 classes from 27 different projects coming from different domains, discarding the trivial ones with cyclomatic complexity lower than 5~\cite{panichella2018automated}.
While this represents already a variegate and large-scale empirical study, replications targeting different types of projects are still desirable.

Threats to \emph{conclusion validity} stem from the relationship between the treatment and the outcome.
To analyze the results of our experiments, we use appropriate statistical tests coupled with sufficient repetitions~\cite{campos2017empirical}.
We rely on the Wilcoxon Rank-Sum Test \cite{conover1999practical} for the statistical significance, and we only discuss the statistically significant results.
Moreover, we estimate the differences of the distributions for the observed metrics relying on the Vargha-Delaney effect size statistic~\cite{vargha2000acritique}.

\section{Conclusions}
\label{sec:conclusion}
This paper introduces \pdmosa, an \emph{adaptive} search-based algorithm that optimizes a secondary objective orthogonal to code coverage without any negative effect on the latter.
We instantiate \pdmosa to the problem of generating tests with lower resource demands, focusing on runtime and heap memory consumption along with seven different coverage criteria.
To avoid the overhead of precise performance measurements, we introduce a set of low-overhead performance proxies that estimate computational demands of tests.
\pdmosa incorporates these proxies into the main search loop, enabling/disabling them as a substitute of the crowding distance depending on whether search stagnation is detected or not.

Our empirical study on 110 Java classes shows that \pdmosa achieves results comparable to \dmosa over seven different coverage criteria.
When reaching a similar level of branch coverage, the test suites produced by \pdmosa are less expensive to run in 72\% (for runtime) and in 70\% (for heap memory consumption) of the CUTs.
In these cases, we observe a decrease of \textasciitilde24\% and \textasciitilde15\% in runtime and heap memory consumption, respectively.
Moreover, we evaluate the fault effectiveness of the generated test suites to avoid counter-effects due to the performance optimization:
we show that \pdmosa achieves similar or higher mutation scores for \textasciitilde75\% of the classes under test.

Based on these results, we plan to investigate different directions for future work:
(i) investigate new proxies, evaluate their individual impact and resolve eventual multicollinearity~\cite{o2007caution},
(ii) horizontally enlarge our study by including further Java classes from different projects and domains,
and (iii) instantiate our adaptive approach to other secondary objectives orthogonal to coverage.

\section*{Acknowledgements}
%
Laaber acknowledges the support of the SNSF through project MINCA (no. 165546).

\footnotesize
\bibliographystyle{IEEEtranN}
\bibliography{main}
\begin{IEEEbiography}[{\includegraphics[width=72pt,height=90pt]{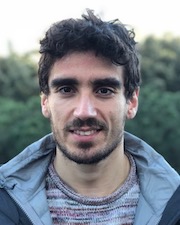}}]{Giovanni Grano}
has been a PhD student since November 2016 at University of Zurich at the software evolution and architecture lab led by Harald Gall.
In his Ph.D he aims at understanding the main factors that affect test quality as well as device novel techniques that allow to automatically generating tests with better design qualities.
His research interests include Search-Based Software Engineering (SBSE), with a main focus on Search-Based Software Testing (SBST), Software Maintenance and Evolution and Empirical Software Engineering. 
Website: \url{https://giograno.me}
\end{IEEEbiography}
\begin{IEEEbiography}[{\includegraphics[width=72pt,height=90pt]{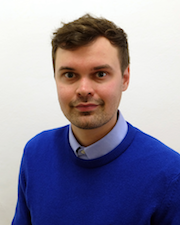}}]{Christoph Laaber}
has been a PhD student since 2016 at University of Zurich at the software evolution and architecture lab led by Harald Gall. Mainly supervised by Philipp Leitner (Chalmers | University of Gothenburg), his research interests are topics at the intersection of software engineering and performance engineering. He mainly works on bringing software performance testing to continuous integration and cloud environments as well as on general topics related to software performance. Website: \url{http://laaber.net}
\end{IEEEbiography}
\begin{IEEEbiography}[{\includegraphics[width=72pt,height=90pt]{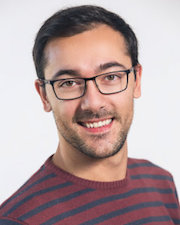}}]{Annibale Panichella} is an Assistant Professor in the Software Engineering Research Group (SERG) at Delft University of Technology (TU Delft) in Netherlands. He is also a research fellow in the Interdisciplinary Centre for Security, Reliability and Trust (SnT), University of Luxembourg.
His research interests include security testing, evolutionary testing, search-based software engineering, textual analysis, and empirical software engineering. He serves and has served as program committee member of various international conference (\textit{e.g.,} ICSE, GECCO, ICST and ICPC) and as reviewer for various international journals (\textit{e.g.,} TSE, TOSEM, TEVC, EMSE, STVR) in the fields of software engineering and evolutionary computation.
 Website: \url{https://apanichella.github.io}
\end{IEEEbiography}
\begin{IEEEbiography}[{\includegraphics[width=72pt,height=90pt]{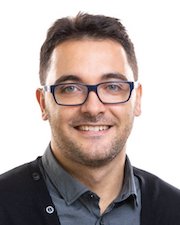}}]{Sebastiano Panichella}is a Computer Science Researcher at Zurich University of Applied Science (ZHAW).  His research interests are in the domain of Software Engineering (SE) and cloud computing (CC).
His research is funded by one SNF Grant and one Innosuisse Project.
He is the author of several papers appeared in International Conferences and Journals.  These research work involved studies with industrial companies and open-source projects and received best paper awards. 
He serves and has served as program committee member of various international conference and as a reviewer for various international journals in the fields of software engineering.
He is Editorial Board Member of Journal of Software: Evolution and Process, Review Board member of the EMSE and TOSEM, and Lead Guest Editor of special issues at EMSE and IST Journals.  
He was selected as one of the top-20 Most Active Early Stage Researchers Worldwide in SE.
Website: \url{https://spanichella.github.io}
\end{IEEEbiography}

\end{document}